\pdfoutput=1

\documentclass[11pt]{article}

\usepackage[preprint]{acl}

\usepackage{times}
\usepackage{latexsym}

\usepackage[T1]{fontenc}

\usepackage[utf8]{inputenc}

\usepackage{microtype}

\usepackage{inconsolata}

\usepackage{graphicx}
\usepackage{subcaption}
\usepackage{xspace}
\usepackage[most]{tcolorbox}
\usepackage{pifont, xcolor}
\usepackage{placeins}
\usepackage{todonotes}
\usepackage[frozencache,cachedir=.]{minted}
\usepackage{listings}

\newcommand{\collex}{\texttt{CollEX}\xspace}

\title{CollEX -- A Multimodal Agentic RAG System\\Enabling Interactive Exploration of  Scientific Collections}

\author{
    \textbf{Florian Schneider}\textsuperscript{$\dagger$}, \textbf{Narges Baba Ahmadi}\textsuperscript{$\dagger$~*}, \textbf{Niloufar Baba Ahmadi}\textsuperscript{$\dagger$~*} \\
    \textbf{Iris Vogel}\textsuperscript{$\ddagger$}, \textbf{Martin Semmann}\textsuperscript{$\dagger$}, \textbf{Chris Biemann}\textsuperscript{$\dagger$}
\\
\\
 \textsuperscript{$\dagger$}Hub of Computing and Data Science \\
 \textsuperscript{$\ddagger$}Center for Sustainable Research Data Management \\
 University of Hamburg, Germany \\
 \small{
    \textbf{Correspondence:} \href{mailto:florian.schneider-1@uni-hamburg.de}{florian.schneider-1@uni-hamburg.de}
 }
\\
 \tiny{
    \textsuperscript{*}Equal contributions, sorted alphabetically.
 }
\\
}

\begin{document}
\maketitle
\begin{abstract}
In this paper, we introduce \collex, an innovative multimodal agentic Retrieval-Augmented Generation (RAG) system designed to enhance interactive exploration of extensive scientific collections.
Given the overwhelming volume and inherent complexity of scientific collections, conventional search systems often lack necessary intuitiveness and interactivity, presenting substantial barriers for learners, educators, and researchers.
\collex addresses these limitations by employing state-of-the-art Large Vision-Language Models (LVLMs) as multimodal agents accessible through an intuitive chat interface.
By abstracting complex interactions via specialized agents equipped with advanced tools, \collex facilitates curiosity-driven exploration, significantly simplifying access to diverse scientific collections and records therein.
Our system integrates textual and visual modalities, supporting educational scenarios that are helpful for teachers, pupils, students, and researchers by fostering independent exploration as well as scientific excitement and curiosity.
Furthermore, \collex serves the research community by discovering interdisciplinary connections and complementing visual data.
We illustrate the effectiveness of our system through a proof-of-concept application containing over 64,000 unique records across 32 collections from a local scientific collection from a public university.
\end{abstract}
\section{Introduction}
\label{sec:intro}
The exploration of scientific knowledge is a cornerstone of human progress.
However, the vast and rapidly growing body of scientific literature presents significant challenges for educators and learners, who often find themselves overwhelmed by the sheer volume and complexity of information.
Despite advancements in information retrieval and knowledge discovery~\cite{santhanam-etal-2022-colbertv2,zhu2023llm4ir,li2024matching}, existing search systems for rich and complex data often lack the interactivity, intuitiveness, and cross-modal search capabilities~\cite{faysse2024colpali,zhai2023siglip,zhao2023mmrag} to engage diverse audiences, such as students, teachers, or researchers.
This limitation negatively affects educational settings where fostering curiosity is essential.

With this paper, we introduce \collex, a multimodal agentic Retrieval-Augmented Generation (RAG) system~\cite{lewis2020rag,zhao-etal-2023-retrieving,xie2024lvlmagents} and reimagine how users explore and interact with scientific collections such as those collected and managed by the Smithsonian Institution\footnote{\url{https://www.si.edu/collections}} or local collections from public universities.
\collex uses state-of-the-art Large Vision-Language Models (LVLMs)\cite{liu2023llava,google2023gemini,openai2024gpt4o,yang2024qwen2.5,google2025gemma} as multimodal agents~\cite{xie2024lvlmagents,wang2024llmagentssurvey} through an intuitive chat interface.
Unlike traditional systems requiring expert knowledge, \collex promotes curiosity-driven exploration, simplifying access and increasing engagement.

\begin{figure}[!t]
    \centering
    \includegraphics[width=1.\linewidth]{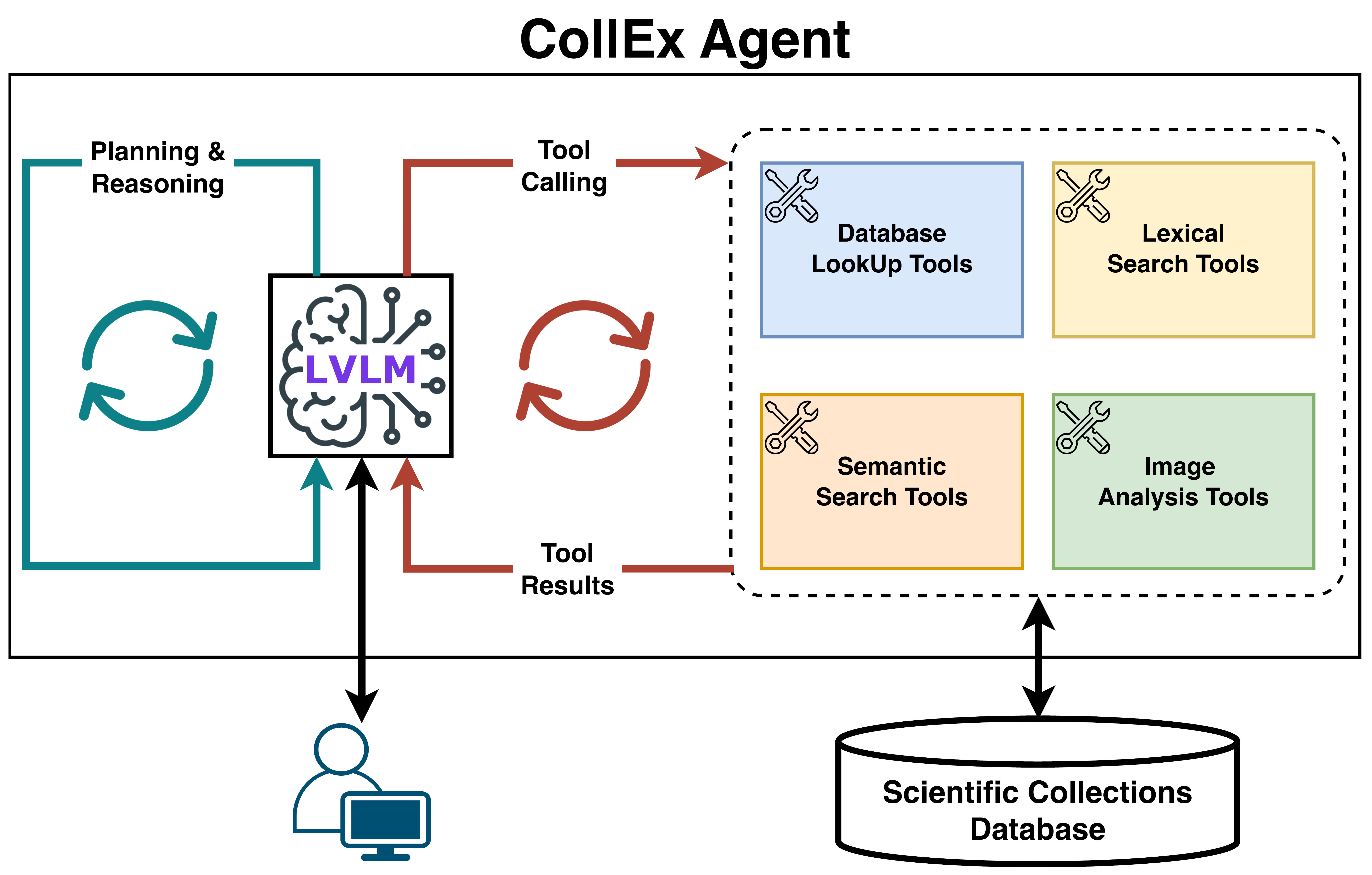}
    \caption{An overview of the \collex Agentic System.}
    \label{fig:agent}
\end{figure}
The core of \collex is its multimodal agentic RAG system, which abstracts complex interactions using specialist agents equipped with various tools~\cite{patil2024gorilla}.
This simplifies the exploration of extensive scientific collections, catering to users with diverse backgrounds and expertise, thereby overcoming accessibility issues~\cite{achiam2014framework}.
The system integrates texts and images, offering intuitive access to scientific concepts.

\collex is especially beneficial in education, fostering curiosity and engagement.
For instance, teachers can get inspiration to prepare visually rich lessons, retrieve relevant information, and facilitate interactive assignments.
Pupils can independently explore the collections, transforming static materials into dynamic learning experiences.
Moreover, \collex supports higher education by encouraging independent exploration and enhancing critical thinking skills.

Beyond education, \collex aids researchers in discovering interdisciplinary connections, eventual related work, or visual data complements.
It autonomously enriches search queries, facilitating easier contextualization and increasing accessibility to scientific collections, thereby supporting national and international scientific connectivity~\cite{weber2018national}.

This paper introduces \collex's general system architecture\footnote{We publish the open-source code here:%
\url{https://github.com/uhh-lt/fundus-murag}}%
and inner workings, combining state-of-the-art LVLMs, advanced prompting and RAG techniques, cross-modal search, and agentic reasoning and planning.

Moreover, we provide three exemplary user stories to demonstrate the system by implementing a proof-of-concept application to explore 32 diverse scientific collections comprising over 64,000 unique items.
\section{Related Work}
\label{sec:related}
\subsection{Cross-Modal Information Retrieval}
Cross-modal information retrieval powered by multimodal embeddings is the key foundation for systems navigating or exploring textual and visual data such as \collex.
Recent developments in multimodal embedding models~\cite{tschannen2025siglip2} that compute semantically rich dense vector representations in an aligned vector space for texts and images, have significantly improved over the popular text-image encoder model, commonly known as CLIP~\cite{radford2021clip}.
This progress was primarily driven by billion-scale high-quality text-image datasets~\cite{schuhmann2022laion5b}, improvements in architecture and training regimes~\cite{zhai2023siglip}, and improved Vision Transformers~\cite{alabdulmohsin2023vitso}
Despite their applications in ``pure'' information retrieval settings, the image encoders of the multimodal embedding models also play a crucial role in the advancement of Large Vision Language Models (LVLMs)~\cite{liu2023llava,yang2024qwen2.5,geigle2025centurio} as they are often used to compute the visual tokens processed by the LVLMs.

\subsection{Multimodal Retrieval Augmented Generation}
Multimodal RAG~\cite{zhao2023mmrag} systems integrate various knowledge formats, including images, code, structured databases, audio, and video, to enhance the knowledge of LVLMs at inference time.
\citet{zhao2023mmrag} further highlight that such multimodal data helps mitigate hallucinations and improve interpretability and reasoning by grounding responses in diverse multimodal information.
\citet{riedler2024beyondtext} demonstrate the advantages of incorporating images into textual retrieval systems within industrial applications.
Their findings suggest that image-derived textual summaries often outperform purely embedding-based multimodal approaches.

\subsection{Agentic RAG}
As described above, traditional RAG systems combine LLMs' or LVLMs' generative capabilities with external knowledge bases to enhance their outputs.
Yet these methods are typically constrained by static workflows and linear processes, restricting their adaptability in complex tasks involving multi-step reasoning and dynamic data quries.
Recently, agentic RAG has emerged as an extension of traditional RAG systems by employing autonomous AI agents in a loop within the RAG pipeline.
Agentic RAG employs agentic design patterns and prompting such as reflection, planning, tool utilization, and multi-agent collaboration, enabling systems to iteratively refine and plan retrieval strategies and adapt dynamically to real-time and context-sensitive queries~\cite{singh2025agentic,xie2024lvlmagents,li-etal-2024-mmedagent}.
For example, \citet{schopf-matthes-2024-nlp} introduced NLP-KG, a system specifically designed for exploratory literature search in NLP.
NLP-KG supports users in exploring unfamiliar NLP fields through semantic search and conversational interfaces grounded in scholarly literature, effectively bridging the gap between exploratory and targeted literature search tasks.
\citet{xie2024lvlmagents} further extends the concept of autonomous LLM agents into the multimodal domain, demonstrating how LVLMs can perceive and interpret diverse data types beyond text, such as images and videos.
Further, they outline critical components necessary for multimodal agent functionality, including visual perception and planning.

With \collex, we integrate a powerful multimodal embedding model for effective cross-modal semantic search with state-of-the-art LVLMs employed as autonomous agents in a multimodal RAG system.
With this, we support educational scenarios by fostering independent exploration, scientific curiosity, and excitement that benefit teachers, pupils, students, and researchers alike.

\section{The \collex System}
\label{sec:arch}
This section describes the \collex system, i.e., its architecture and core components, as well as the data to be explored.
\subsection{\collex Data}
\label{sec:arch:data}
\begin{figure}[!ht]
    \centering
    \includegraphics[width=1.\linewidth]{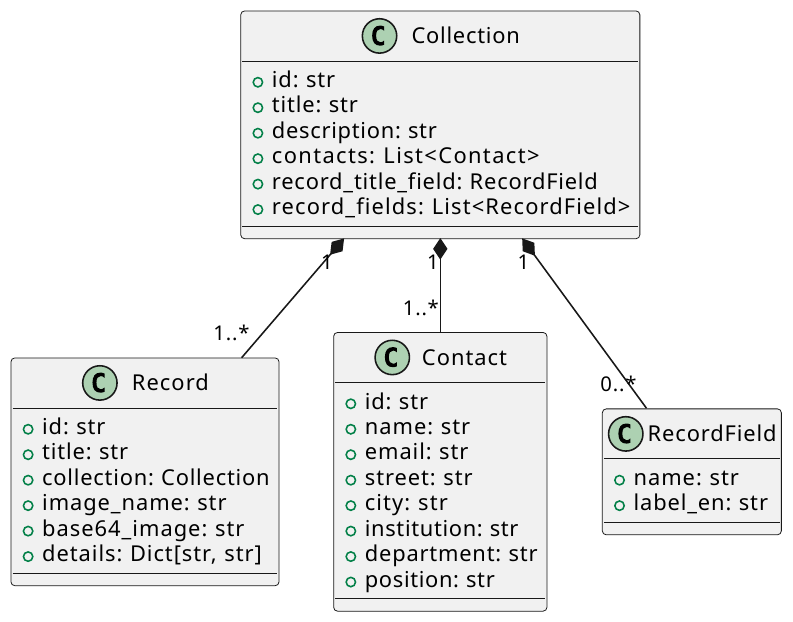}
    \caption{The \collex Data Schema}
    \label{fig:arch:data:schema}
\end{figure}

Since \collex is a multimodal agentic RAG system, to understand the system, it is essential to know the data it operates on.
\paragraph{Schema.} We provide the simplified data schema as a UML class diagram in Figure~\ref{fig:arch:data:schema}.
As the name \collex suggests, our system assists in exploring scientific collections represented by the \texttt{Collection} class.
Each collection has a title, a description, and a list of contacts who own or manage the collection.
More importantly, each collection comprises multiple \texttt{Record}s, which are described by a title, an image, and additional details.
The records' details are described by different \texttt{RecordField}s, depending on the parent collection.

Further, we store embeddings of the collection titles and descriptions as well as the record titles and images computed by a SigLIP~\cite{zhai2023siglip} model\footnote{\href{https://huggingface.co/google/siglip-so400m-patch14-384}{siglip-so400m-patch14-384}} in the vector database.

\begin{figure}[!ht]
    \centering
    \begin{subfigure}{0.45\linewidth}
        \includegraphics[width=\textwidth]{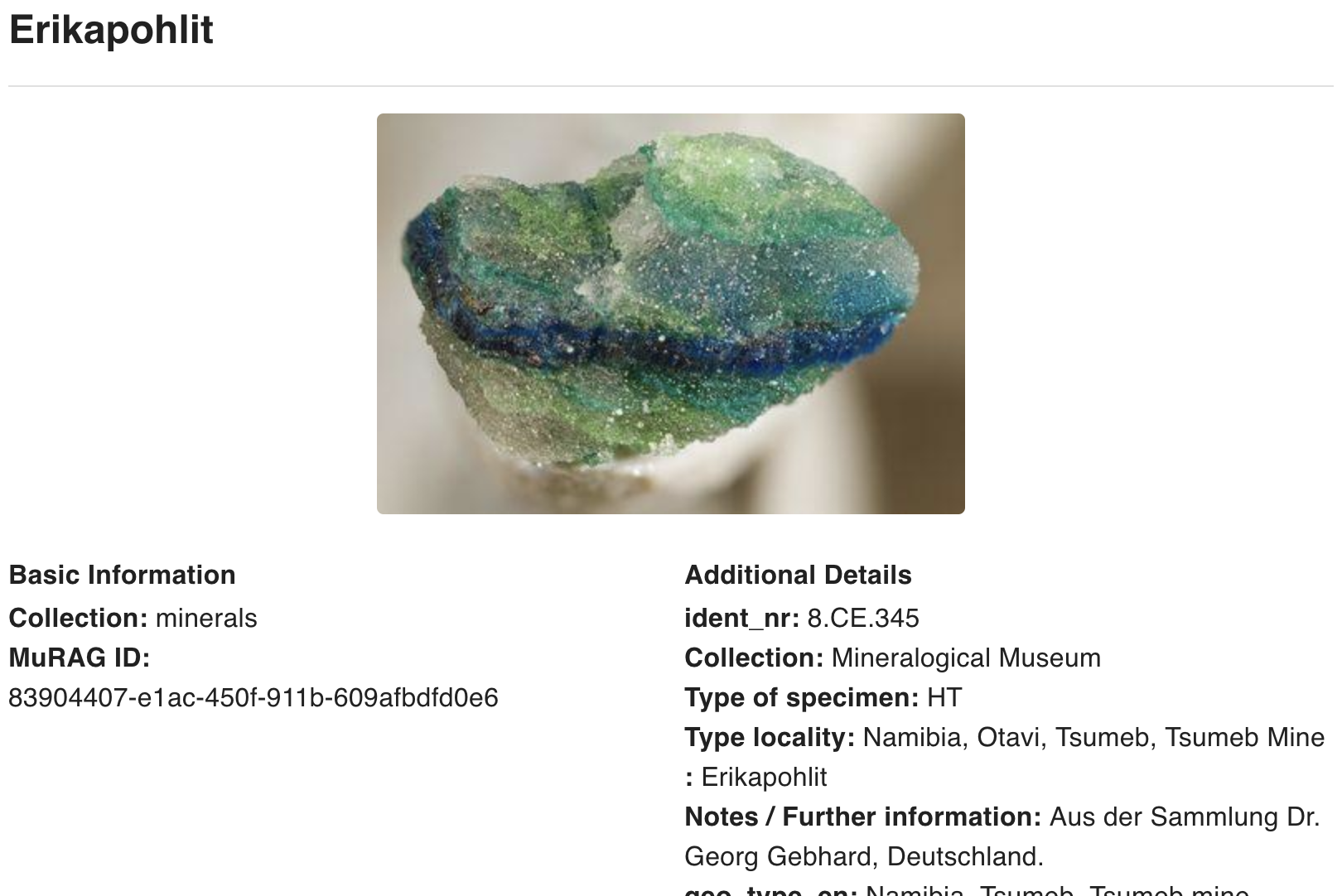}
        \caption{Example Record 1}
    \end{subfigure}
    \hfill
    \begin{subfigure}{0.45\linewidth}
        \includegraphics[width=\textwidth]{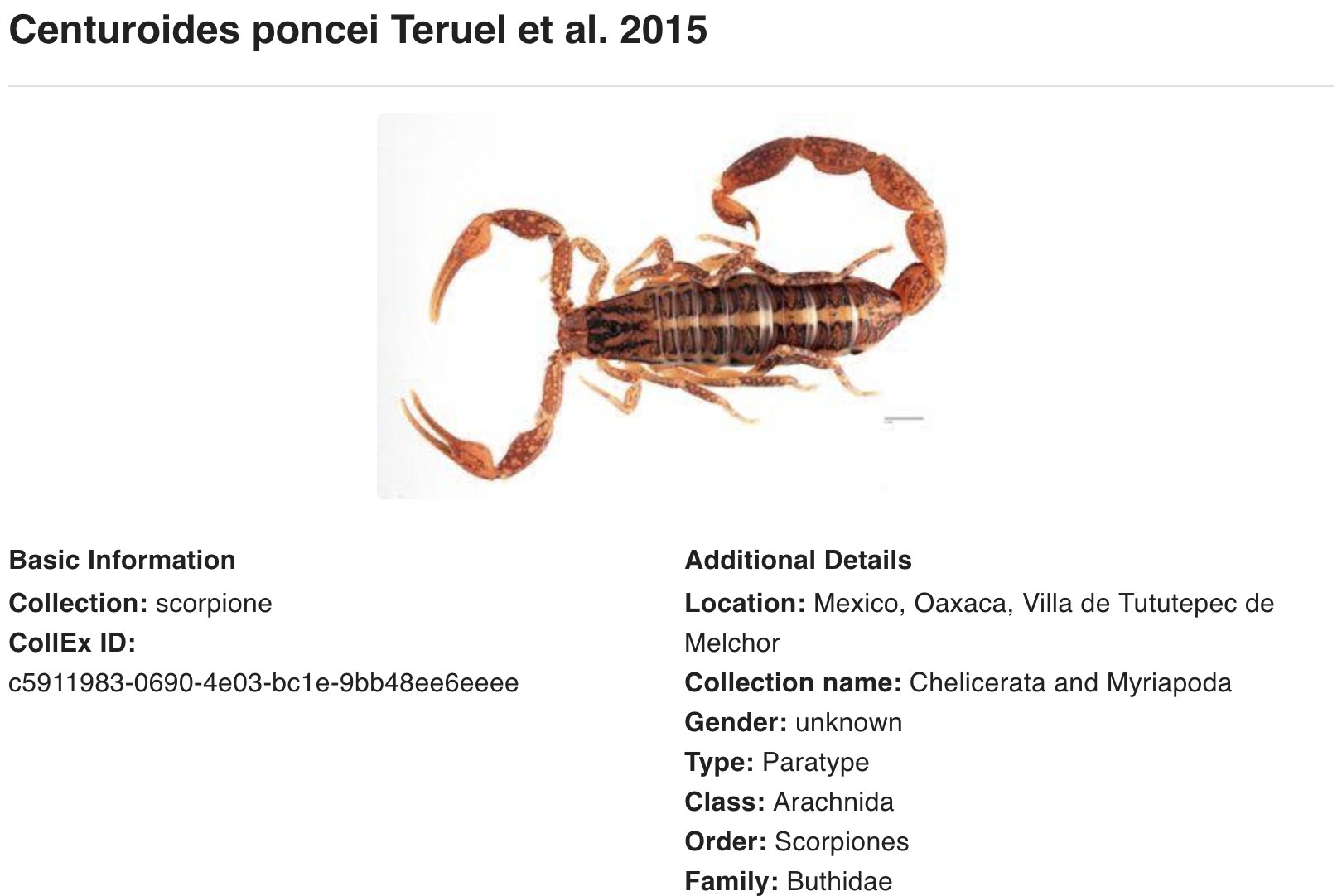}
        \caption{Example Record 2}
    \end{subfigure}
    
    \begin{subfigure}{0.45\linewidth}
        \includegraphics[width=\textwidth]{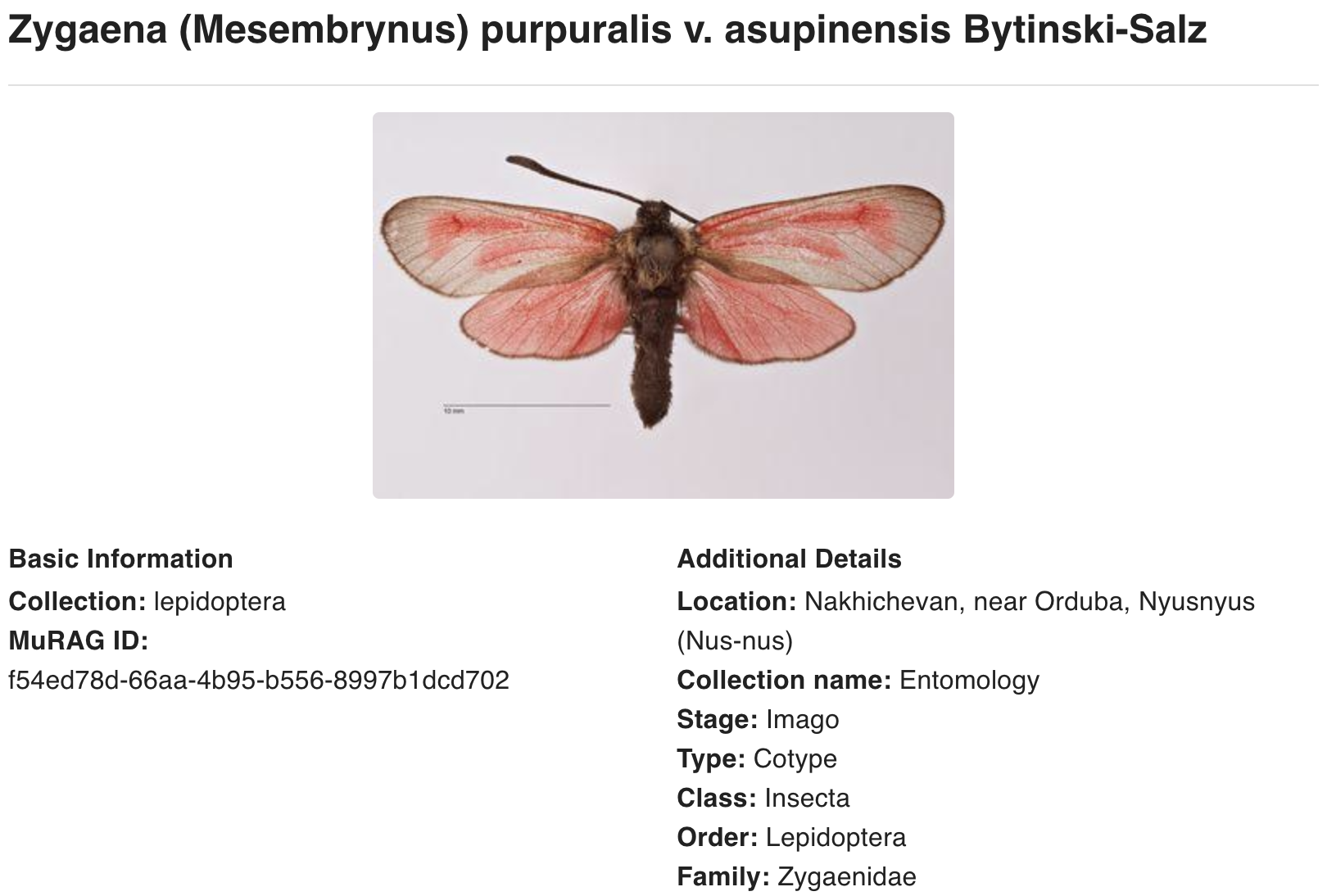}
        \caption{Example Record 3}
    \end{subfigure}
    \hfill
    \begin{subfigure}{0.45\linewidth}
        \includegraphics[width=\textwidth]{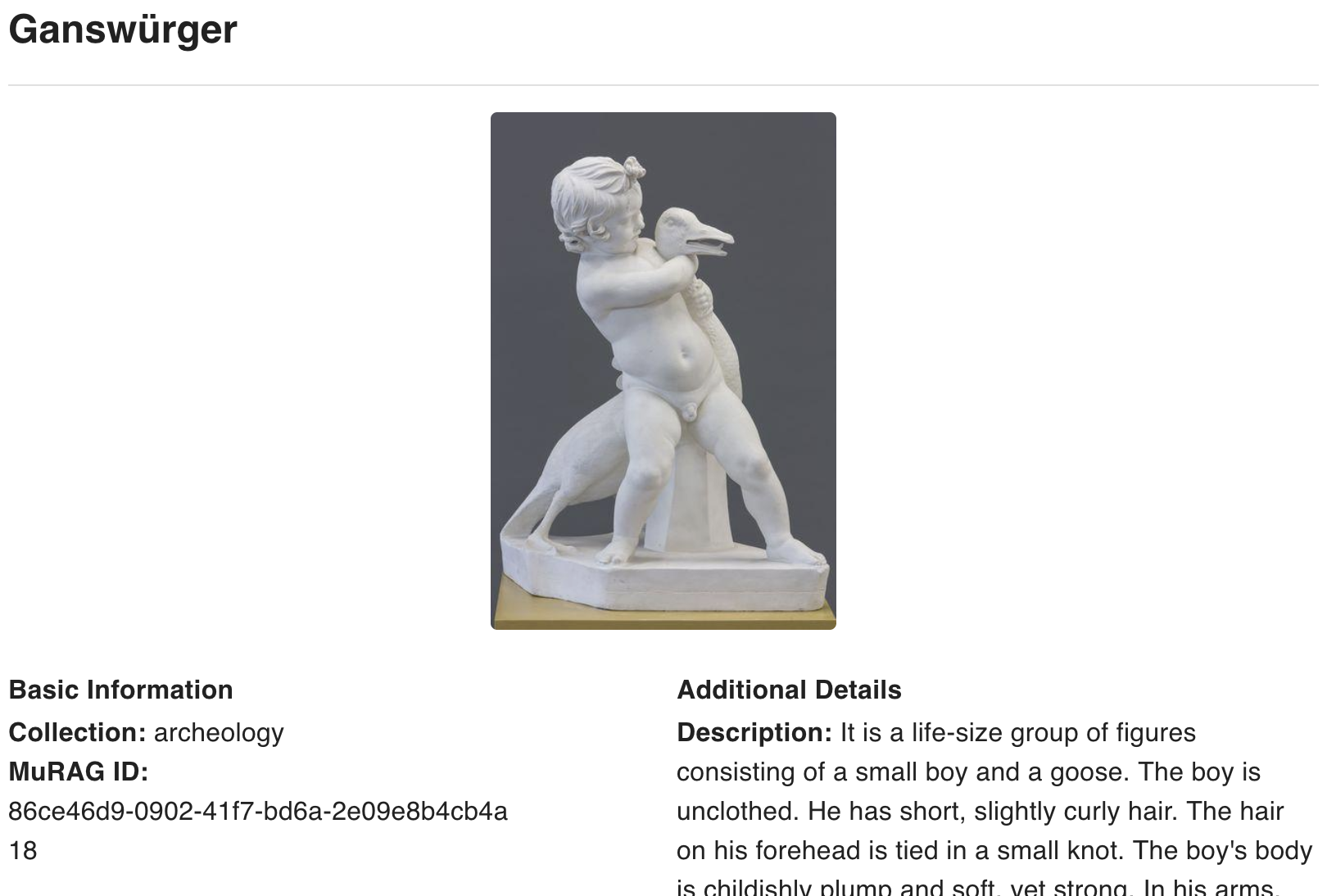}
        \caption{Example Record 4}
    \end{subfigure}
    
    \caption{Examples records contained in the \collex database.}
    \label{fig:arch:data:examples}
\end{figure}

\paragraph{Examples.} To get a better idea of the data, we provide four example records in Figure~\ref{fig:arch:data:examples}.

In total, in our \collex proof-of-concept application, we store 64,469 unique records in 32 collections.

\subsection{\collex System Architecture}
\label{sec:arch:arch}
\begin{figure}[!ht]
    \centering
    \includegraphics[width=1.\linewidth]{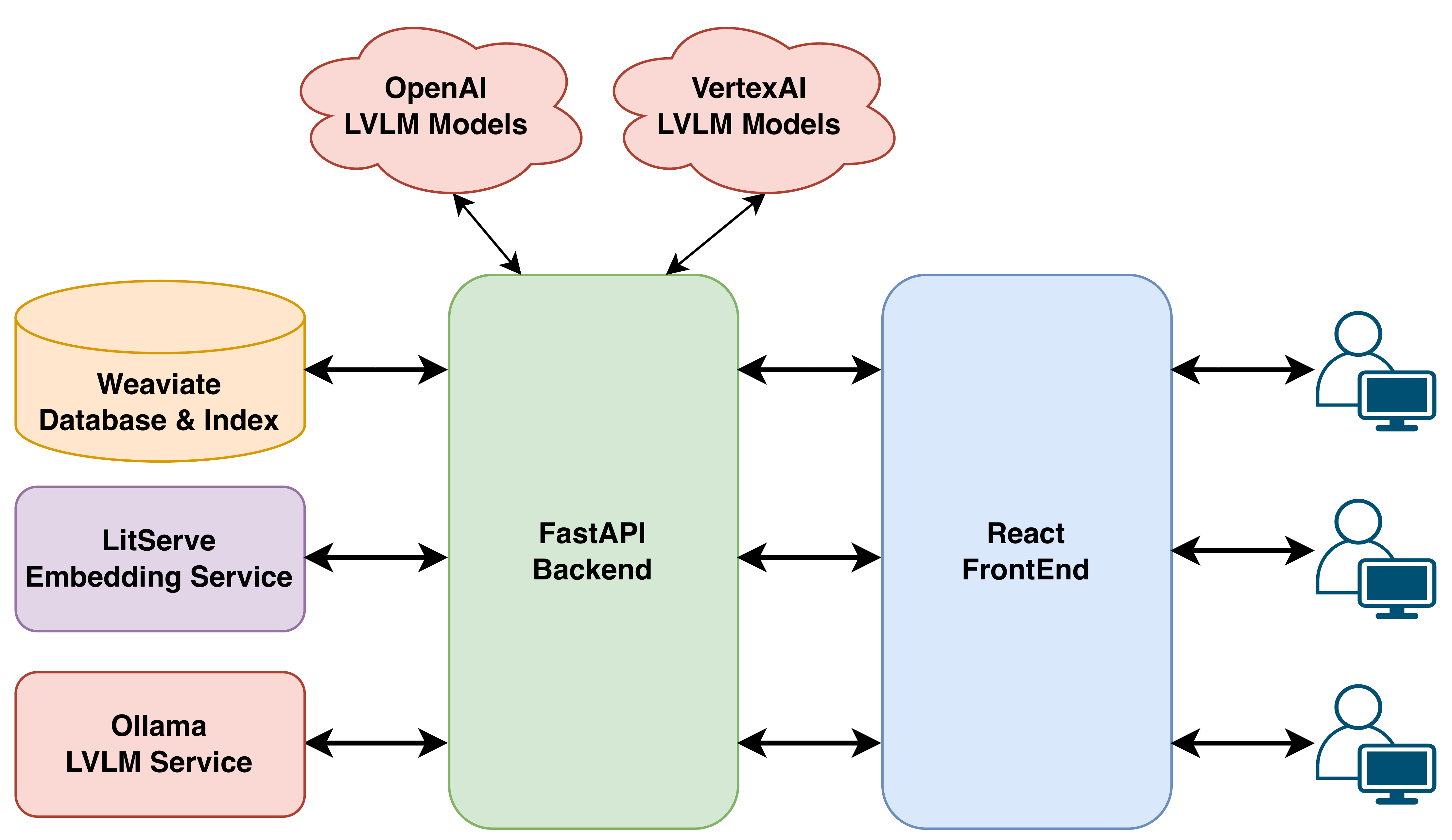}
    \caption{Overview of the \collex system architecture.}
    \label{fig:arch}
\end{figure}

\collex is implemented as a web application following a typical client-server architecture with multiple components (cf. Figure~\ref{fig:arch}), which are described in the following.
Each component is containerized using Docker\footnote{\url{https://www.docker.com}}, and the whole system is deployed using Docker Compose\footnote{\url{https://docs.docker.com/compose/}}.
\paragraph{Backend:} This component is the core of \collex responsible for orchestrating and communicating between the other components.
Its functionality is implemented by several services, e.g., to retrieve information from the database, embed user queries, manage chat sessions of different users, or communicate with LVLMs hosted by different providers.
Most importantly, it implements the \collex Agent described in Section~\ref{sec:arch:agent}.
Its core functionality is exposed as REST API endpoints implemented using \textit{FastAPI}\footnote{\url{https://fastapi.tiangolo.com/}}.
\paragraph{Database:} We store all data using \textit{weaviate}\footnote{\url{https://weaviate.io/}}.
More specifically, we precomputed all text and image embeddings (cf. \S\ref{sec:arch:data}) and store them in an HNSW~\cite{malkov2018hnsw} index for efficient semantic search.
Further, to enable lexical search, we store collection descriptions and titles, as well as record titles in a BM25~\cite{robertson2009bm25} index.
Other data, e.g., contacts for collections, are simply stored in the (NoSQL) database without indexing.
\paragraph{Embedding Service:} To efficiently embed user queries of arbitrary texts and images for cross-modal semantic search, we use \textit{LitServe}\footnote{\url{https://lightning.ai/litserve}}.
That is, we serve the same \textit{SigLIP} embedding model used to compute the embeddings stored in the HNSW index and expose the functionality through a REST API.
\paragraph{LVLM Models:} At the core of \collex, we employ a Large Vision-Language Model (LVLM) that handles user queries and powers the agent (cf. \S\ref{sec:arch:agent}).
To (qualitatively) test the effectiveness of different models and not force or restrict users with different privacy constraints, we implemented \collex LVLM-agnostic.
That is, we provide multiple proprietary as well as open-weight LVLMs such as \textit{Gemma3}~\cite{google2025gemma}, \textit{Gemini}~\cite{google2023gemini} 1.5 and 2.0 models, \textit{GPT-4o}~\cite{openai2024gpt4o}, or \textit{o1}~\cite{openai2024o1} to power our multimodal agentic RAG system.
However, one important constraint to the LVLMs is that it must support function calling~\cite{patil2024gorilla}.
\paragraph{Frontend:} We implemented the \collex web application, employing a modern \textit{Vite}\footnote{\url{https://vite.dev/}} + \textit{React Typescript}\footnote{\url{https://react.dev/}} + \textit{Material UI}\footnote{\url{https://mui.com/}} web stack that facilitates a responsive and intuitive user interface.
Futher, the frontend manages user interactions, rendering visualizations, and handles asynchronous requests and responses to ensure a seamless user experience.
\subsection{\collex Agent}
\label{sec:arch:agent}
The \collex agent (cf. Figure~\ref{fig:agent} sits at the core of our multimodal agentic RAG system and is described in the following.

To act as a tool calling agent, we designed an effective prompt for the respective LVLM combining prompt engineering techniques such as (Auto) Chain-of-Thought~\cite{wei2022cot,zhang2023autocot} and ReAct~\cite{zheng2024react,sahoo2024promptsurvey}.
The full prompt is provided in Appendix \ref{appendix:agent_prompt}.
Further, we implement an agentic loop (cf. Listing~\ref{lst:agentic_loop}, which gets executed for each user request.
\begin{listing}[!ht]
    \centering
    \begin{minted}[fontsize=\footnotesize,breaklines]{python}
def run_agentic_loop(user_request, chat_history):
    # Add the user's message to the chat history.
    chat_history.append(user_request)
    
    # Step 1: Generate initial response using the updated chat history.
    lvlm_response = generate_response(chat_history)
    update_chat_history(lvlm_response, chat_history)
    
    # Step 2: Loop while the response contains tool call instructions.
    while is_tool_call_response(response):
        # Execute tool calls and obtain the resulting tool messages.
        tool_responses = execute_tool_calls(response)
        
        # Update the chat history with the tool responses.
        update_chat_history(tool_responses, chat_history)
        
        # Generate a new response with the updated chat history.
        lvlm_response = generate_response(chat_history)
        update_chat_history(lvlm_response, chat_history)
    
    # Step 3: Extract and return the final message content.
    message = get_message_content(lvlm_response)
    return message
\end{minted}
    \caption{Pseudo code of the agentic loop implemented for the \collex agent.}
    \label{lst:agentic_loop}
\end{listing}
By executing this loop, we enable iterative planning, reasoning, and tool calling of the LVLM, i.e., the agent.
Note that the user requests, as well as the tool responses, can be arbitrarily interleaved text-image messages.
In each iteration, the agent reasons whether it needs to invoke one of the following tools to fulfill the user’s request satisfactorily.
\paragraph{DataBase Lookup Tool:} This tool provides a comprehensive interface for querying the \collex database.
It allows the agent to retrieve aggregate statistics, get records and collections by unique identifiers, or list all collections.

\paragraph{Lexical Search Tool:} This tool enables textual searches over the collections and records in the database by querying the BM25 index through \textit{weaviate}.

\paragraph{Similarity Search Tool:} This tool allows for efficient semantic similarity search to find relevant records or collections.
It supports both textual and image-based cross-modal or uni-modal similarity searches by querying the HNSW index through \textit{weaviate}.
Further, we employ query-rewriting techniques~\cite{ma-etal-2023-query} to enhance the original user request and improve the search results.

\paragraph{Image Analysis Tool:} This tool offers advanced image processing capabilities tailored for images of the records.
It includes functions to generate descriptive captions, answer questions about the visual content, extract textual content from the images, or detect objects within images, which is useful for extracting interesting details about recorded images.
We implemented this functionality by employing an LVLM with task-specific prompts (cf. Appendix~\ref{appendix:image_prompts}).
%

%


%
\section{System Demonstration}
\label{sec:demo}
In the following, we demonstrate \collex showcasing some general functionality and two exemplary user stories depicted by screenshots of the app\footnote{The screenshots were taken in an earlier version of the app, which we named ``FUNDus!'' assistant. This name originated from the name of the original database but was replaced by \collex in later versions for a more general name.}.
Due to the limited space to display the screenshots and the thereby induced readability issues because of the small image sizes, we provide high-resolution screenshots in Appendix~\ref{appendix:user_stories}.
\subsection{General Functionality}
\begin{figure*}[!ht]
    \centering
    \begin{subfigure}{0.32\linewidth}
        \includegraphics[width=\textwidth]{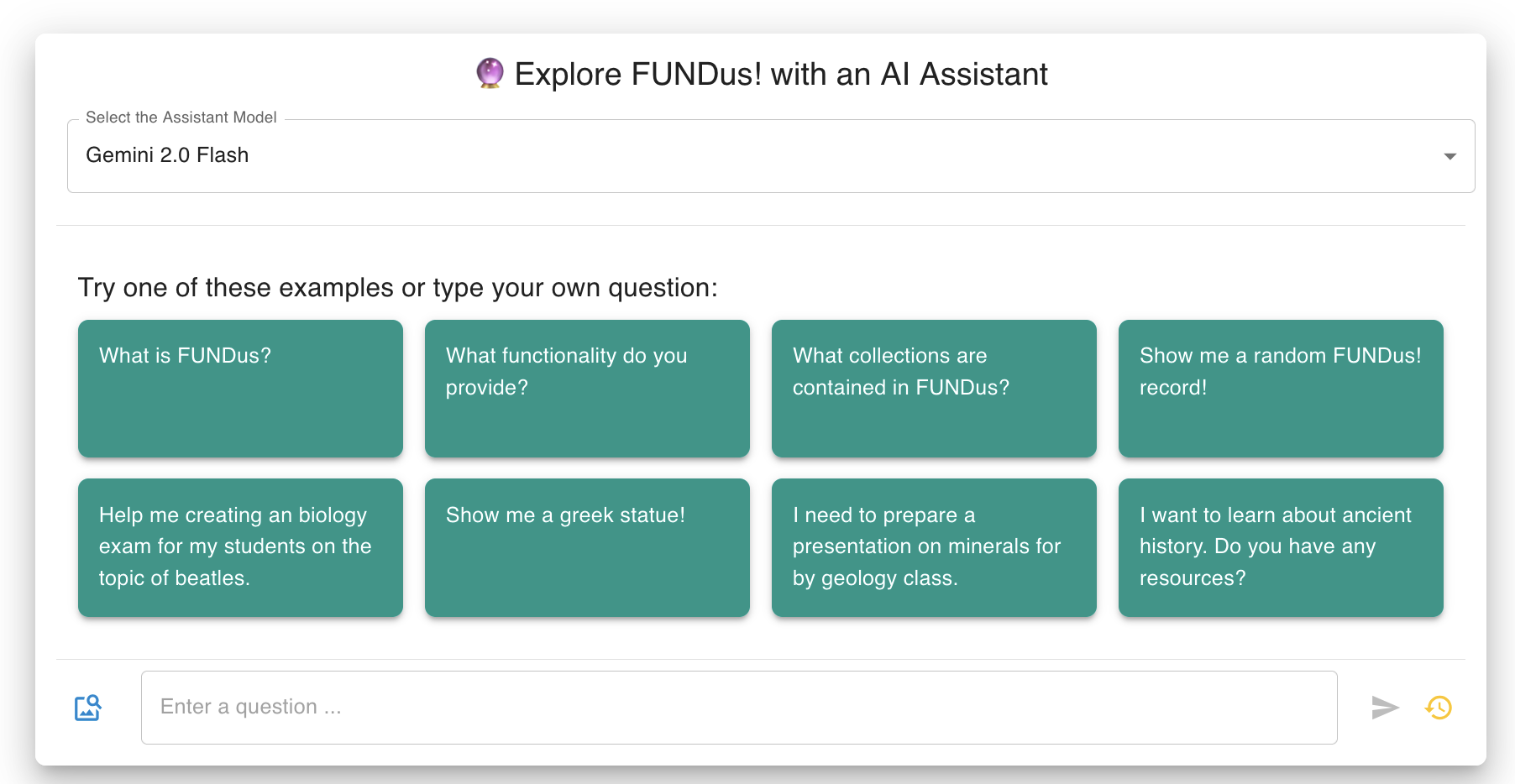}
        \caption{The \collex start page.}
        \label{fig:demo:general:1}
    \end{subfigure}
    \hfill
    \begin{subfigure}{0.32\linewidth}
        \includegraphics[width=\textwidth]{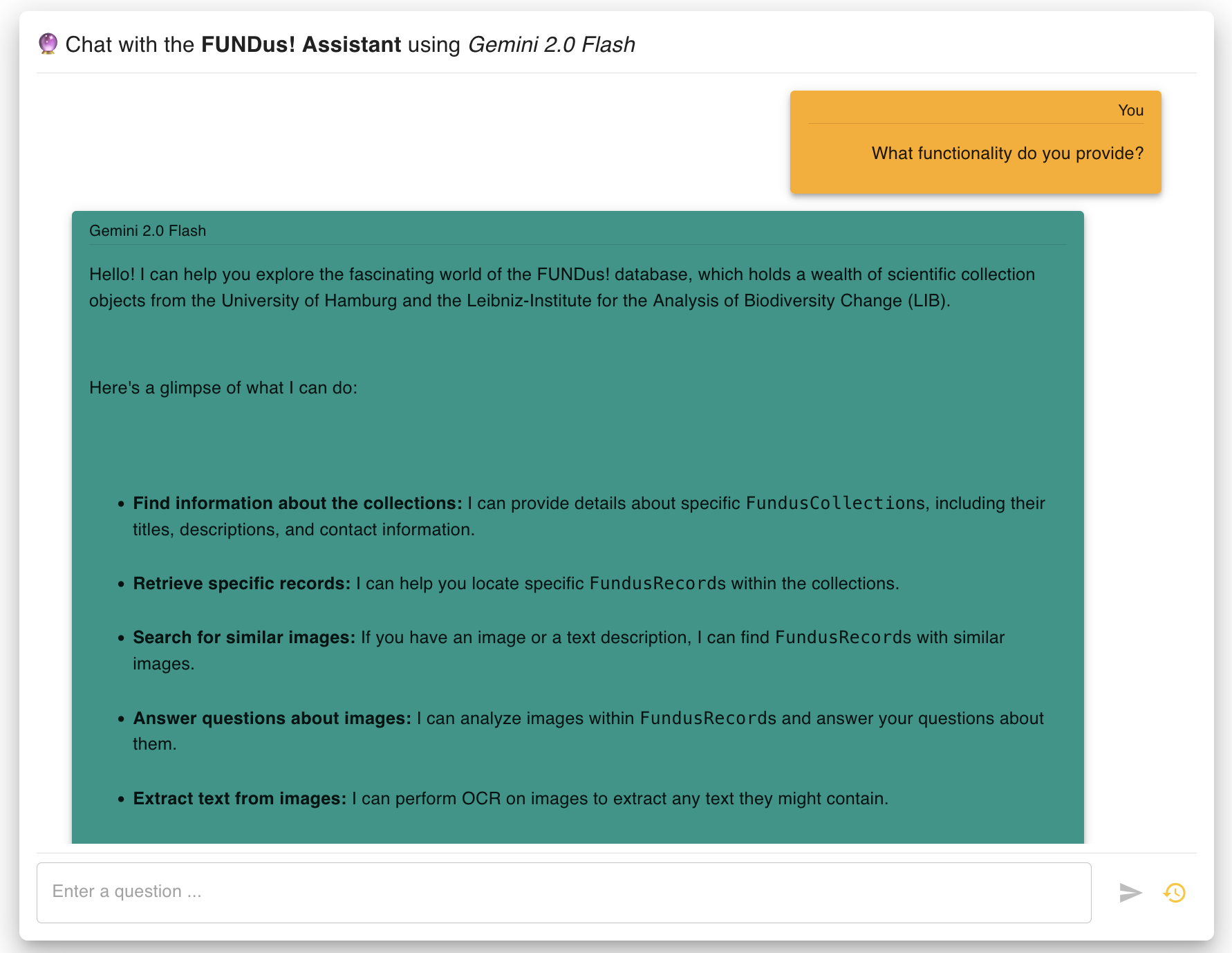}
        \caption{General functionality query.}
        \label{fig:demo:general:2}
    \end{subfigure}
    \hfill
    \begin{subfigure}{0.32\linewidth}
        \includegraphics[width=\textwidth]{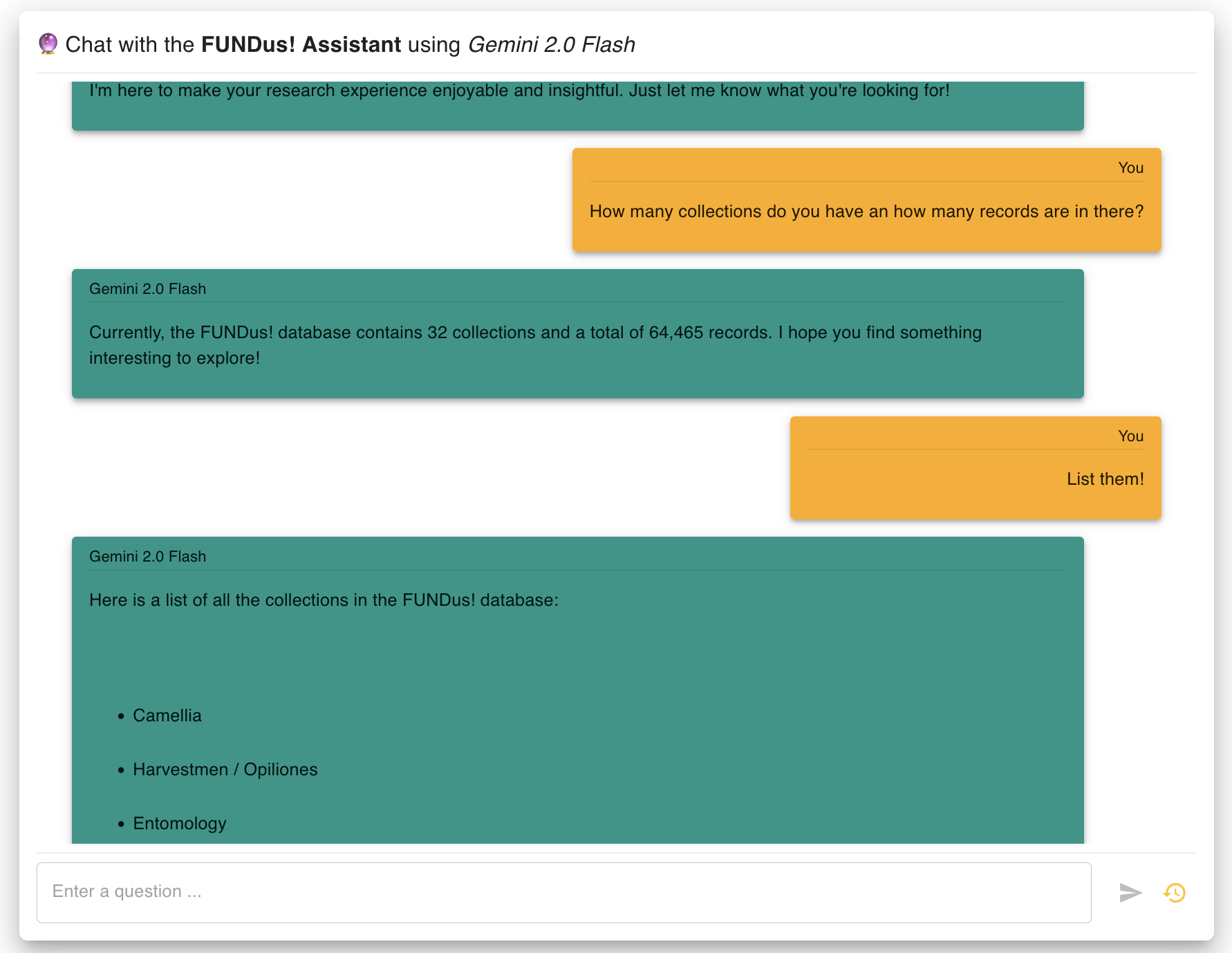}
        \caption{Records and Collections query.}
        \label{fig:demo:general:3}
    \end{subfigure}
    
    \caption{Show-casing \collex general functionality.}
    \label{fig:demo:general}
\end{figure*}

In this demonstration, we present some of the general functionality of \collex in Figure~\ref{fig:demo:general} (or Figure~\ref{fig:demo:highres:general} for high-resolution screenshots).

When a user opens the app in her browser, she sees the start page (cf. Figure \ref{fig:demo:general:1}).
On this page, she can pick the LVLM that powers the system for the chat session she is about to start.
Further, she can click on one of the example prompts to kick-start her \collex experience and get an idea of what the system is capable of.
If she is not interested in trying one of the examples, she can enter an individual question or any arbitrary request in the text input field.

For our example, she picked one of the examples asking the \collex agent about its general functionality.
The agent's responses are always rendered in markdown, and in this case, the answer contains ``a glimpse of what'' the agent can do (cf. Figure \ref{fig:demo:general:2}).

Next, she asks for statistics about the number of records and collections in the database and finally lets the agent explicitly list the collections (cf. Figure~\ref{fig:demo:general:3}).
In the backend, the LVLM makes multiple calls to the \textit{Database Lookup Tool} and prints the received results in a human-readable way.

\subsection{Geology Class Presentation}
\begin{figure*}[!ht]
    \centering
    \begin{subfigure}{0.32\linewidth}
        \includegraphics[width=\textwidth]{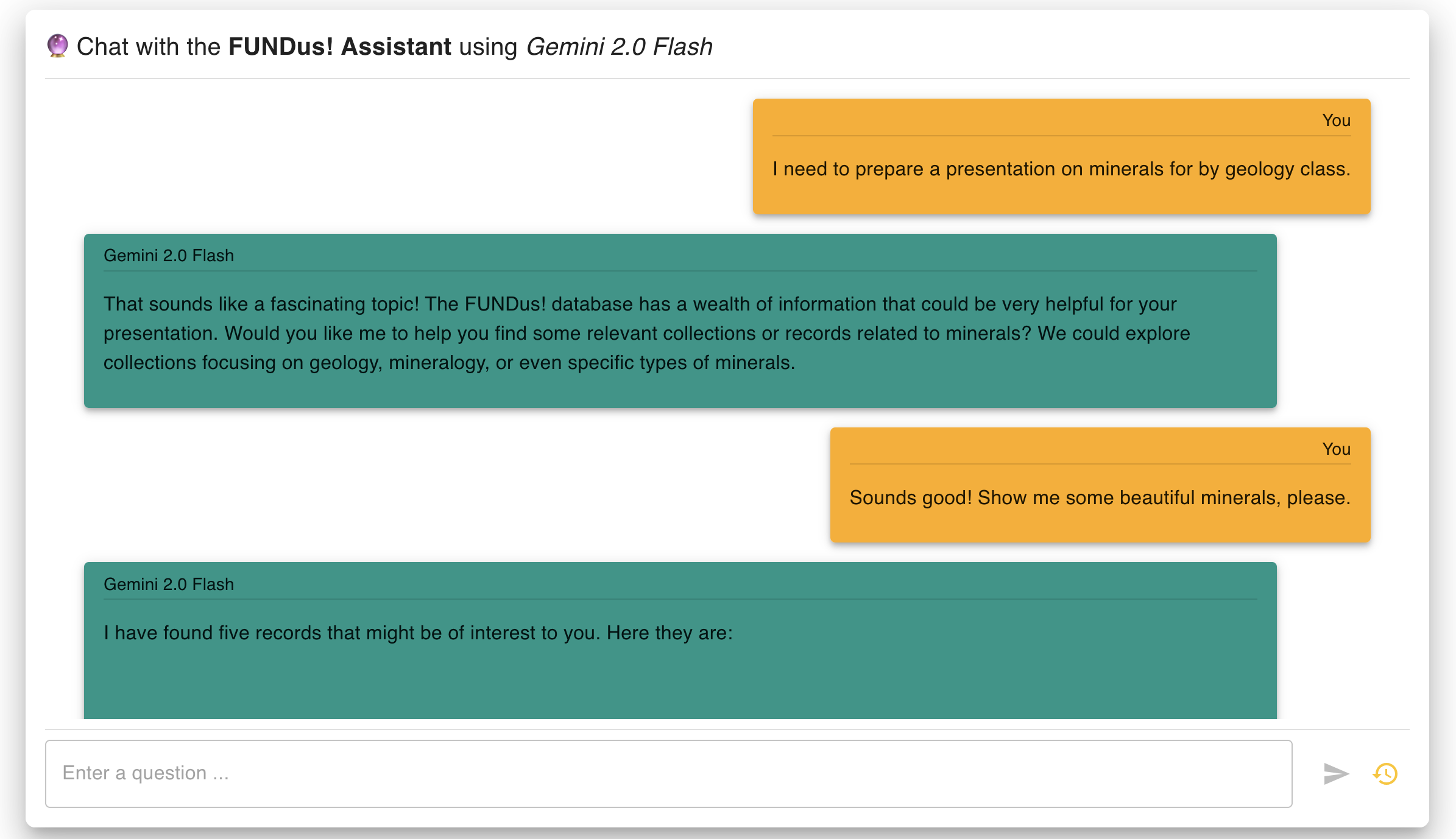}
        \caption{Start of the chat.}
        \label{fig:demo:geo_class:1}
    \end{subfigure}
    \hfill
    \begin{subfigure}{0.32\linewidth}
        \includegraphics[width=\textwidth]{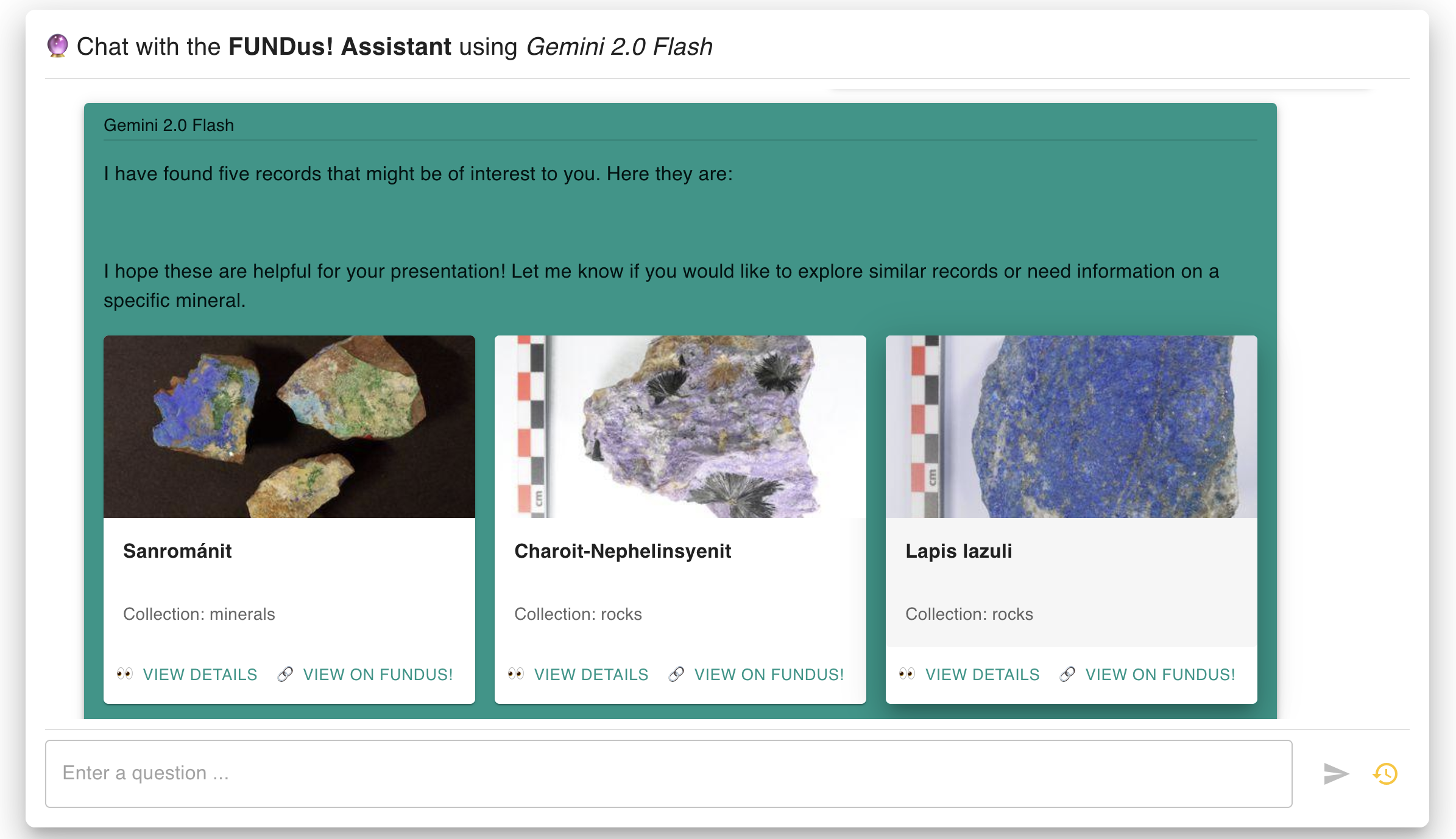}
        \caption{Search results for the user query.}
        \label{fig:demo:geo_class:2}
    \end{subfigure}
    \hfill
    \begin{subfigure}{0.32\linewidth}
        \includegraphics[width=\textwidth]{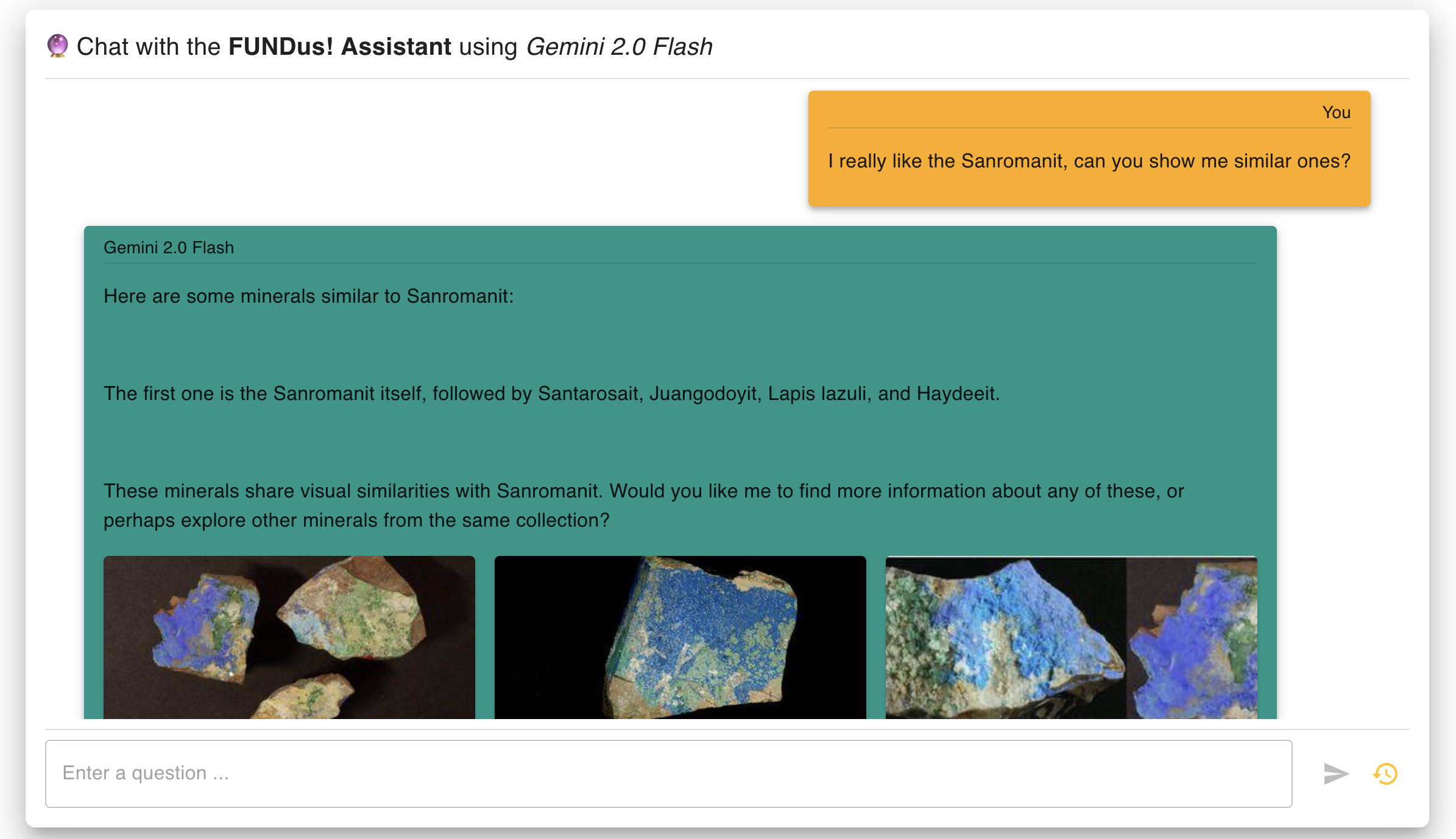}
        \caption{Image similarity search results.}
        \label{fig:demo:geo_class:3}
    \end{subfigure}

    \begin{subfigure}{0.32\linewidth}
        \includegraphics[width=\textwidth]{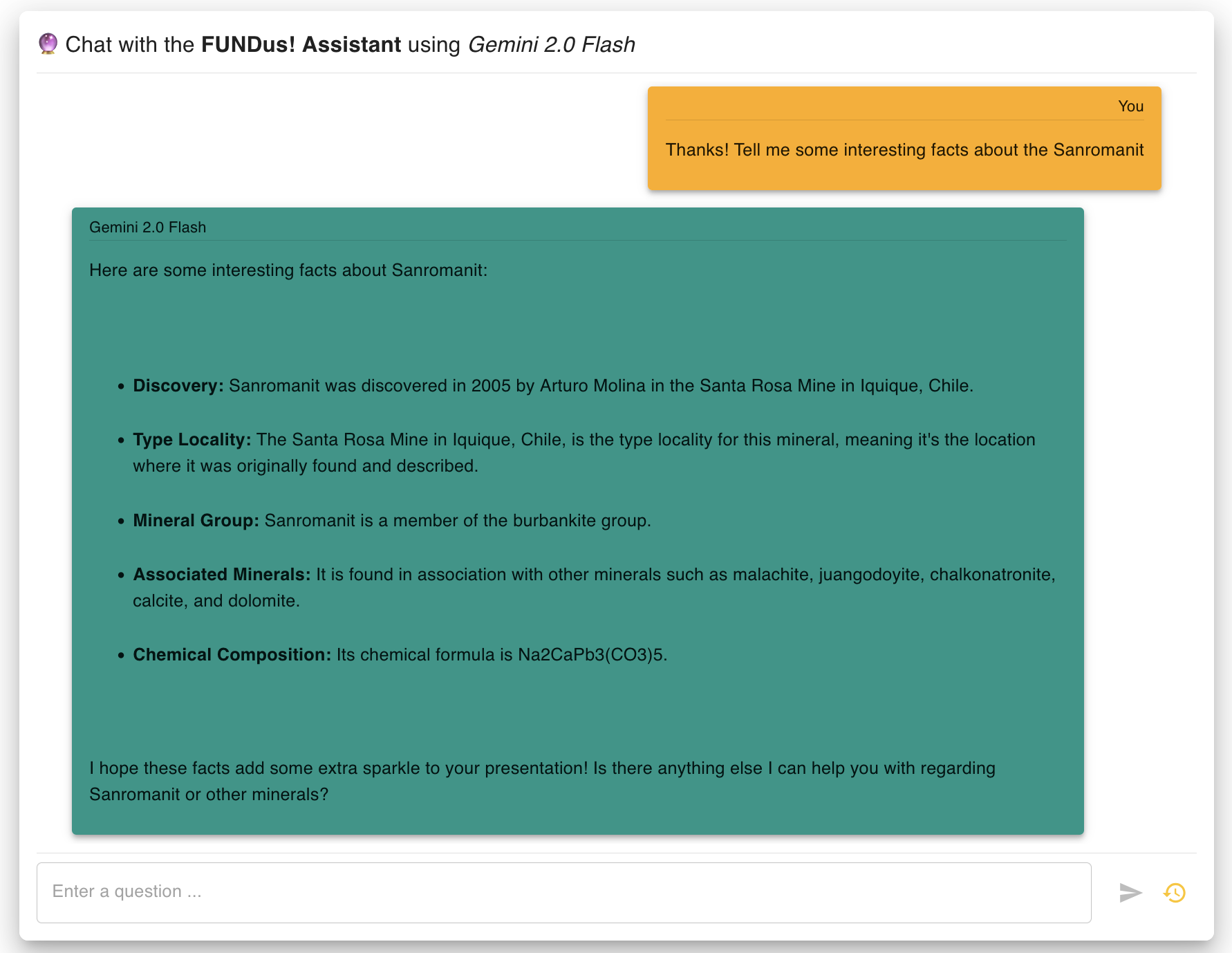}
        \caption{Requesting more details.}
        \label{fig:demo:geo_class:4}
    \end{subfigure}
    \hfill
    \begin{subfigure}{0.32\linewidth}
        \includegraphics[width=\textwidth]{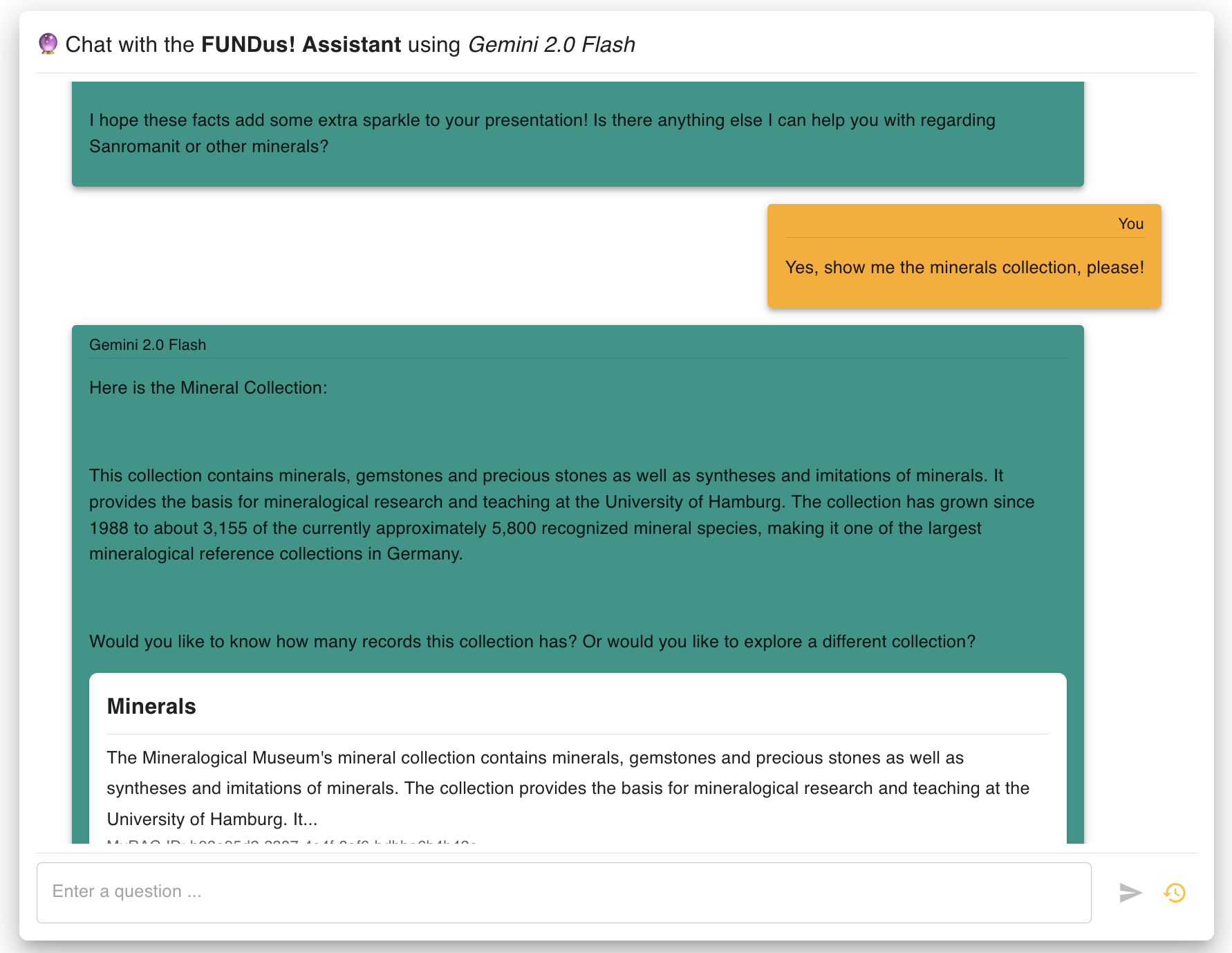}
        \caption{Showing the minerals collection.}
        \label{fig:demo:geo_class:5}
    \end{subfigure}
    \hfill
    \begin{subfigure}{0.32\linewidth}
        \includegraphics[width=\textwidth]{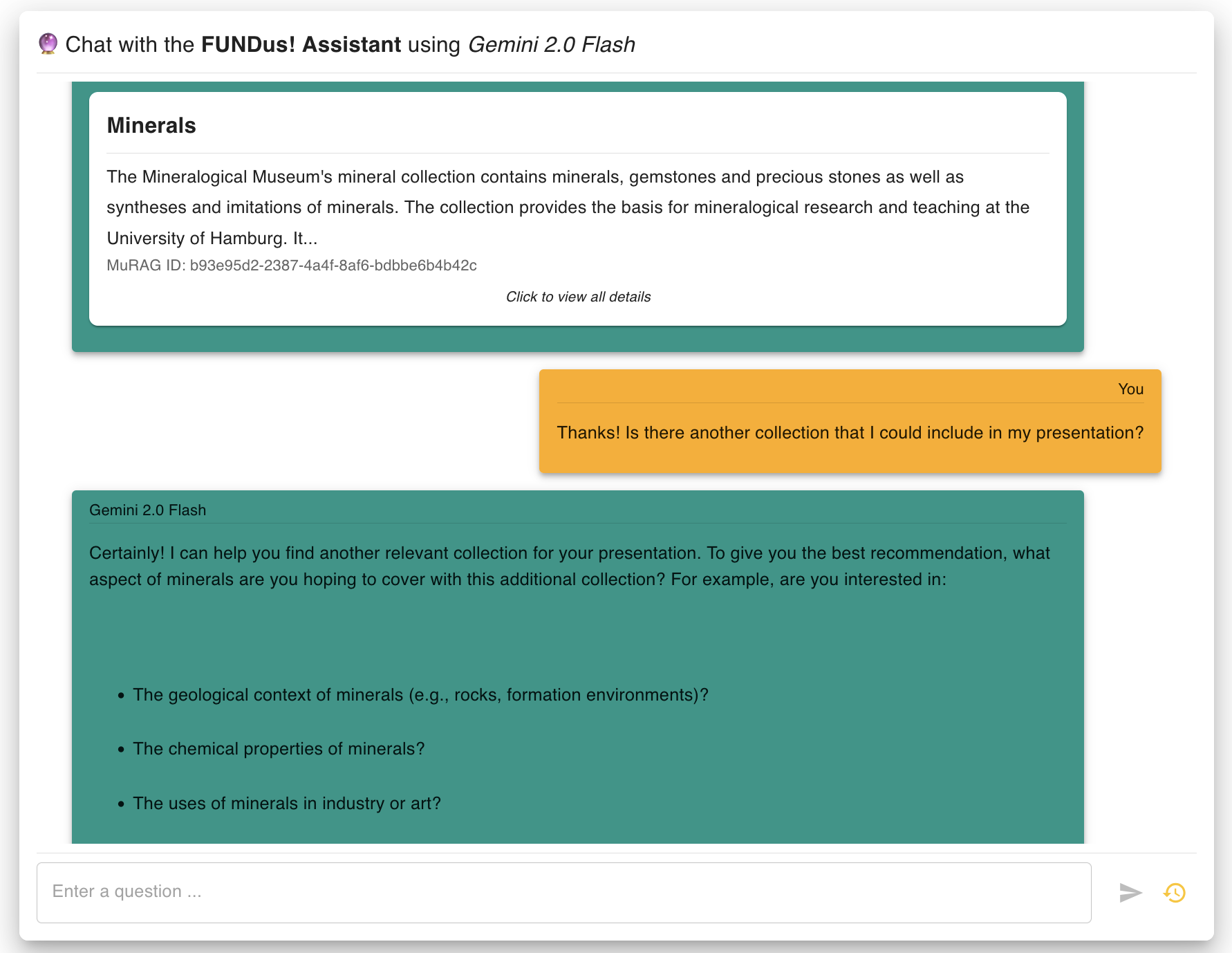}
        \caption{Follow-up query.}
        \label{fig:demo:geo_class:6}
    \end{subfigure}
    
    \caption{A demonstration of \collex based on an exemplary use case of getting inspiration for a geology class presentation.}
    \label{fig:demo:geo_class}
\end{figure*}

In this user story (cf. Figure~\ref{fig:demo:geo_class} or~\ref{fig:demo:highres:geo_class}), Alice needs inspiration for a presentation she has to create about her geology class.

She starts the chat by telling the assistant what her goal is, and the assistant provides her with some ideas on how to find interesting material (cf. Figure~\ref{fig:demo:geo_class:1}).

She likes the suggestions and asks the agent to show her some beautiful minerals.
In the backend, by executing the agentic loop (cf. Listing \ref{lst:agentic_loop}), the LVLM reasons about how to best fulfill the user request and decides to use the text-to-image similarity search provided by the \textit{Similarity Search Tool} with an initial query ``beautiful minerals''.
The specialized query-rewriter agent expands the query to ``a photo of beautiful minerals, geology'', which is then sent to the embedding service to compute the embedding used for the ANN search on the record image embedding vector index.
This returns a list of the top-k best matching records as JSONs as the tool response fed back to the \collex agent.
The decides to return the retrieved records in the form of special rendering tags as instructed (cf. the prompt in Appendix~\ref{appendix:agent_prompt}) in addition to a user-friendly message.
The frontend creates and generates custom rending components to display the records to the user (cf. Figure~\ref{fig:demo:geo_class:2}).

Alice especially likes the first mineral, a ``Sanrománit'', and asks the agent to find similar-looking minerals (cf. Figure~\ref{fig:demo:geo_class:3}).
This triggers the image-to-image similarity search.
After the agentic loop, the backend sends the model's response, including the special rendering tags, to the front end, which displays it to the user.

Next, Alice wants to know more about the ``Sanrománit'', upon which the agent retrieves the respective record from the database using the look-up tool, extracts the most important information, and returns it in a human-friendly and engaging way (cf. Figure~\ref{fig:demo:geo_class:4}).

The user wants to get more general information about the mineral collection, which in turn is presented to her using another special rendering tag for collections (cf. Figures~\ref{fig:demo:geo_class:5} and \ref{fig:demo:geo_class:6}).

Finally, Alice asks about other collections from which she could get inspiration for her presentation.
Since this is an ambiguous query, the agent asks for clarification (cf. Figure~\ref{fig:demo:geo_class:6}).

\subsection{Finding an Exhibition Piece}
\begin{figure*}[!ht]
    \centering
    \begin{subfigure}{0.32\linewidth}
        \includegraphics[width=\textwidth]{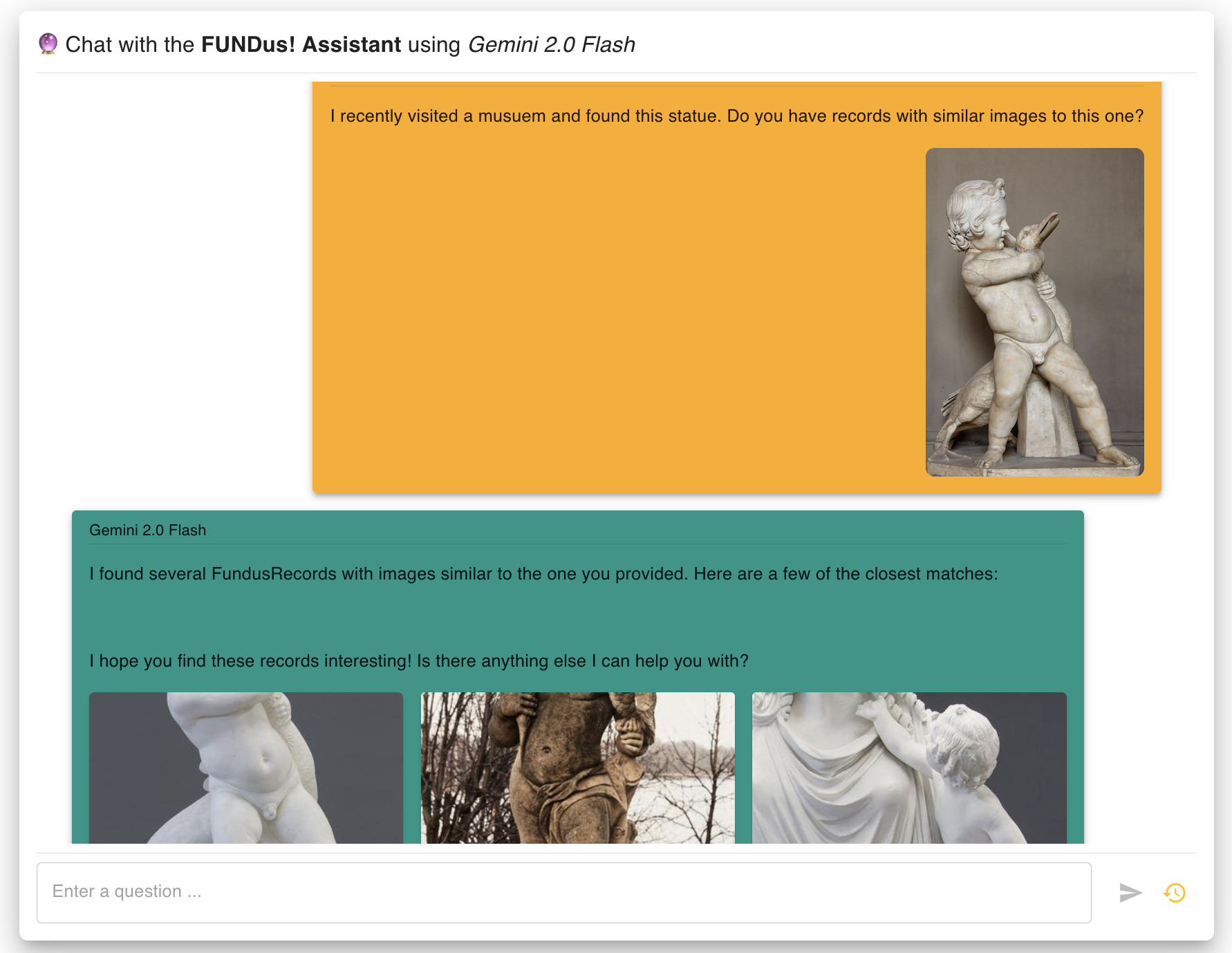}
        \caption{Text-image search request and results.}
        \label{fig:demo:sim_ex:1}
    \end{subfigure}
    \hfill
    \begin{subfigure}{0.32\linewidth}
        \includegraphics[width=\textwidth]{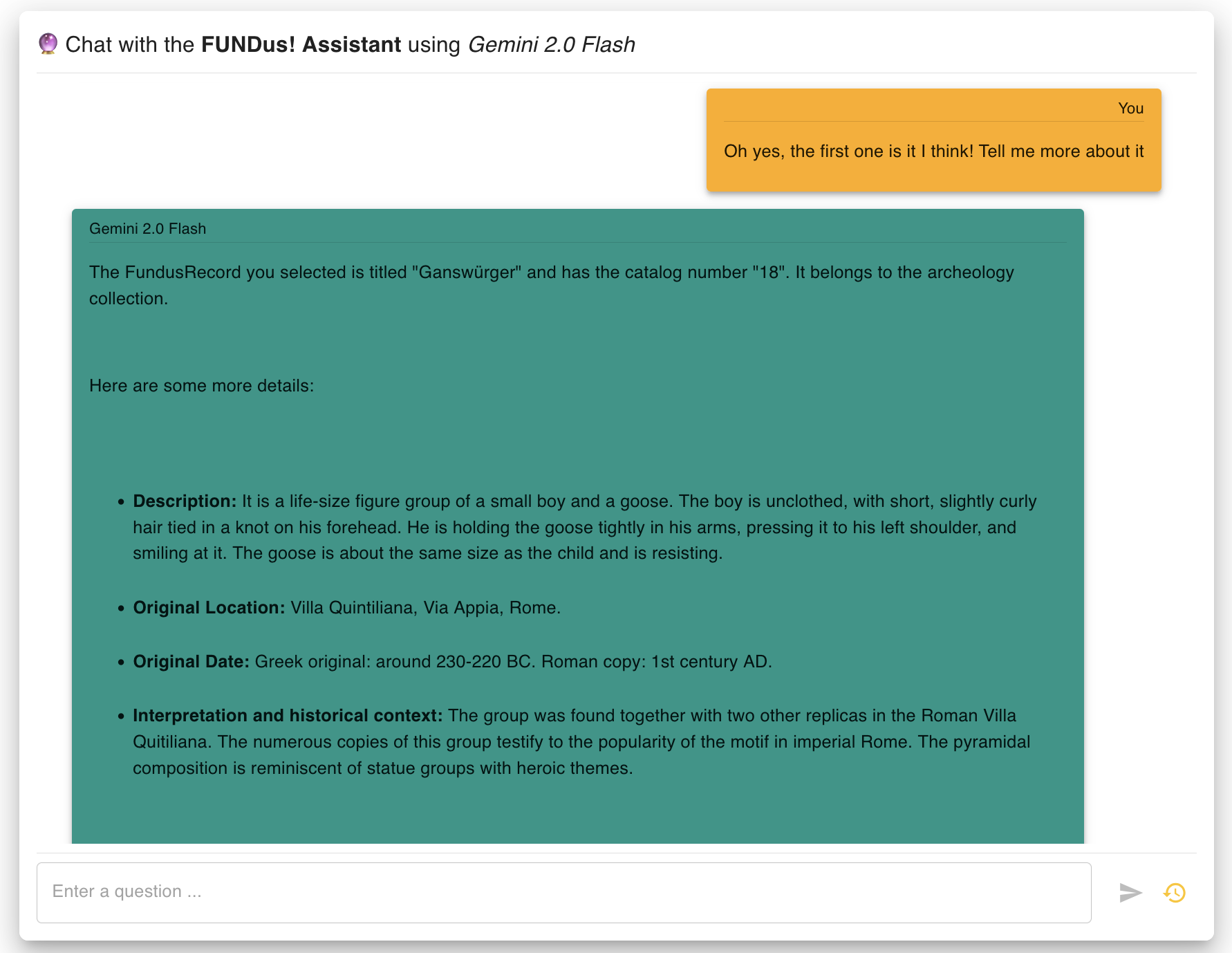}
        \caption{Follow-up details query.}
        \label{fig:demo:sim_ex:2}
    \end{subfigure}
    \hfill
    \begin{subfigure}{0.32\linewidth}
        \includegraphics[width=\textwidth]{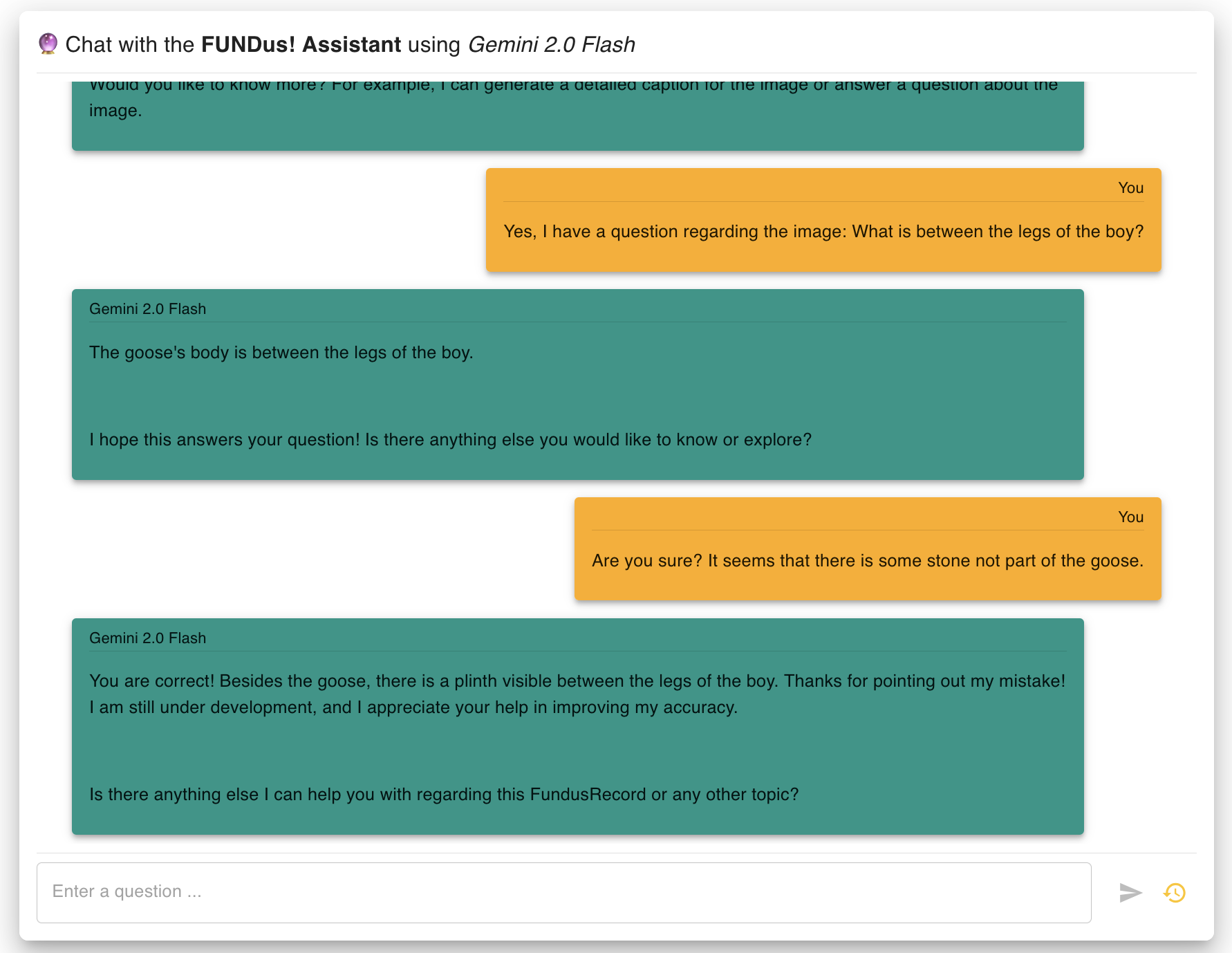}
        \caption{Image analysis queries.}
        \label{fig:demo:sim_ex:3}
    \end{subfigure}

    \caption{A demonstration of \collex based on an exemplary use case of finding an exhibition piece.}
    \label{fig:demo:sim_ex}
\end{figure*}

In this user story (cf. Figure~\ref{fig:demo:sim_ex}), a user, Bob, recently visited a museum and took a photo of an interesting statue.

However, he forgot to take notes and decides to use the \collex assistant to get more information (cf. Figure~\ref{fig:demo:sim_ex:1})
In the backend, this triggers the image-to-image similarity search and returns the best-matching records, which are displayed to the user by special rendering tags. 

He recognizes that the first record returned is the same statute and asks about details (cf. Figure~\ref{fig:demo:sim_ex:2}).

Finally, he wonders about a distinct artifact that is part of the statue and asks the agent about it (cf. Figure~\ref{fig:demo:sim_ex:3}).
This triggers a call to the visual question answering (VQA) functionality of the \textit{Image Analysis Tool}, which returns an answer.
Bob is not convinced by that first answer and asks the agent to analyze the image again.
This triggers another call to the VQA tool as well as to the image captioning tool.
Finally, combining the tool results, the agent correctly identifies the unknown artifact as a plinth of the goose statue (cf. Figure~\ref{fig:demo:sim_ex:3}).
\section{Conclusion}
\label{sec:conclusion}
In this work, we introduced \collex, an innovative multimodal agentic RAG system aimed at facilitating interactive and intuitive exploration of extensive scientific collections.
Leveraging state-of-the-art LVLMs, \collex provides a powerful yet user-friendly interface for diverse audiences, such as pupils, students, educators, or researchers.
Our proof-of-concept implementation, covering over 64,000 scientific items across 32 diverse collections, successfully demonstrates the system's potential, showcasing capabilities such as cross-modal search, advanced semantic retrieval, and agent-driven interactions.
Additionally, \collex serves as a versatile blueprint that can be straightforwardly applied to other scientific collections.

In conclusion, with \collex, we presented an innovative system to interactively explore scientific collections, enhancing educational and research-oriented applications, thereby positively contributing to the broader scientific community.
%
%

\bibliography{stripped}
\section{Limitations}
\label{sec:limitations}
Despite the promising potential of our introduced system, we acknowledge several limitations summarized in the following:

Firstly, user experience when using \collex heavily depends on the capabilities of the underlying LVLMs.
If a model misinterprets the user intent, invokes incorrect or irrelevant tools, misuses parameters, misunderstands tool responses, or fails to communicate results clearly and engagingly, the application's usability and user satisfaction significantly suffers.
Such issues might lead to frustration among users, diminishing their excitement in the tool and thereby scientific exploration which is the opposite of our intention.

Secondly, \collex performs optimally with proprietary LVLMs, which can create dependency and privacy issues including substantial ongoing costs and reliance on external model providers.
Although the system supports integration with open-source LVLMs, the overall user experience often suffers, as open-source alternatives generally lag behind in accuracy, responsiveness, and general robustness.

Thirdly, \collex currently integrates an extensive range of tools that, while offering powerful capabilities, sometimes overwhelms or confuses the LVLM.
This complexity can lead to inappropriate or inefficient tool use, further impacting the overall user experience negatively.
A potential solution would involve reorganizing the system from a single agent into multiple specialized agents managed hierarchically by an orchestrator agent.
This would simplify decision-making processes and tool invocation more effectively.
However, since we currently do not rely on any agentic frameworks or libraries to implement \collex, this introduces several challenges such as optimizing the inter-communication between the agents.

Lastly, the current implementation of \collex lacks formal evaluation of both the overall system and its individual components.
This is primarily due to the considerable investment in computational and human resources required for comprehensive user studies and empirical assessments.
Without systematic evaluations, it remains challenging to quantify the true effectiveness, usability, and scalability of the system in real-world contexts.
Therefore, conducting extensive evaluations to validate the system's performance and identify areas for improvement is a priority for future work.
\appendix
\onecolumn
\section{\collex Agent System Instruction}
\label{appendix:agent_prompt}
\begin{tcolorbox}[
    enhanced, 
    breakable,
    skin first=enhanced,
    skin middle=enhanced,
    skin last=enhanced,
]
\begin{minted}[fontsize=\footnotesize,breaklines]{markdown}
# Your Role

You are a helpful and friendly AI assistant that that supports and motivates users as they explore the FUNDus! database.

# Your Task

You will provide users with information about the FUNDus! Database and help them navigate and explore the data.
You will also assist users in retrieving information about specific FundusRecords and FundusCollections.
Your goal is to provide and motivate users with a pleasant and informative experience while interacting with the FUNDus! Database.

# Basic Information about FUNDus!

'''
FUNDus! is the research portal of the University of <REDACTED>, with which we make the scientific collection objects of the University of <REDACTED> and the Leibniz-Institute for the Analysis of Biodiversity Change (LIB) generally accessible. In addition werden provide information about the collections of the Staats- and Universitätsbiliothek <REDACTED>. We want to promote the joy of research! Our thematically arranged offer is therefore aimed at all those who want to use every opportunity for research and discovery with enthusiasm and joy."
There are over 13 million objects in 37 scientific collections at the University of <REDACTED> and the LIB - from A for anatomy to Z for zoology. Some of the objects are hundreds or even thousands of years old, others were created only a few decades ago."

Since autumn 2018, interesting new collection objects have been regularly published here. In the coming months you can discover many of them for the first time on this portal.

We are very pleased to welcome you here and cordially invite you to continue discovering the interesting, exciting and sometimes even bizarre objects in the future. In the name of all our employees who have implemented this project together, we wish you lots of fun in your research and discovery!
'''

# Important Datatypes

In this task, you will work with the following data types:

**FundusCollection**
A `FundusCollection` represents a collection of `FundusRecord`s with details such as a unique identifier,
    title, and description.

    Attributes:
        murag_id (str): Unique identifier for the collection in the VectorDB.
        collection_name (str): Unique identifier for the collection.
        title (str): Title of the collection in English.
        title_de (str): Title of the collection in German.
        description (str): Description of the collection in English.
        description_de (str): Description of the collection in German.
        contacts (list[FundusCollectionContact]): A list of contact persons for the collection.
        title_fields (list[str]): A list of fields that are used as titles for the `FundusRecord` in the collection.
        fields (list[FundusRecordField]): A list of fields for the `FundusRecord`s in the collection.

**FundusRecord**
A `FundusRecord` represents an record in the FUNDus collection, with details such as catalog number,
    associated collection, image name, and metadata.

    Attributes:
        murag_id (int): A unique identifier for the `FundusRecord` in the VectorDB.
        title (str): The title of the `FundusRecord`.
        fundus_id (int): An identifier for the `FundusRecord`. If a `FundusRecord` has multiple images, the records share the `fundus_id`.
        catalogno (str): The catalog number associated with the `FundusRecord`.
        collection_name (str): The unique name of the `FundusCollection` to which this `FundusRecord` belongs.
        image_name (str): The name of the image file associated with the `FundusRecord`.
        details (dict[str, str]): Additional metadata for the `FundusRecord`.

# Tool Calling Guidelines

- Use the available tools whenever you need them to answer a user's query. You can also call multiple tools sequentially if answering a user's query involves multiple steps.
- Never makeup names or IDs to call a tool. If you require information about a name or an ID, use one of your tools to look it up!.
- If the user's query is not clear or ambiguous, ask the user for clarification before proceeding.
- Pay special attention to the fact that you exactly copy and correctly use the parameters and their types when calling a tool.
- If a tool call caused an error due to erroneous parameters, try to correct the parameters and call the tool again.
- If a tool call caused an error not due to erroneous parameters, do not call the tool again. Instead, respond with the error that occurred and output nothing else.

# User Interaction Guidelines

- If the user's request is not clear or ambiguous, ask the user for clarification before proceeding.
- Present your output in a human-readable format by using Markdown.
- To show a FundusRecord to the user, use `<FundusRecord murag_id='...' />` and replace `'...'` with the actual `murag_id` from the record. Do not output anything else. The tag will present all important information, including the image of the record.
- If you want to render multiple FundusRecords, use the tag multiple times in a single line separated by spaces.
- To show a FundusCollection, use `<FundusCollection murag_id='...' />` and replace `'...'` with the actual `murag_id` from the collection. Do not output anything else. The tag will present all important information about the collection.
- If you want to render multiple FundusCollections, use the tag multiple times in a single line separated by spaces.
- Avoid technical details and jargon when communicating with the user. Provide clear and concise information in a friendly and engaging manner.
- Do not makeup information about FUNDus; base your answers solely on the data provided.
\end{minted}
\end{tcolorbox}

\newpage
\section{Query Rewriting System Instructions}
\label{appendix:query_prompts}
In the following, we provide the system instructions for query rewriting functionality used for semantic similarity searches.

\subsection{Text-to-Image Similarity Search}
\begin{tcolorbox}[
    enhanced, 
    breakable,
    skin first=enhanced,
    skin middle=enhanced,
    skin last=enhanced,
]
\begin{minted}[fontsize=\footnotesize,breaklines]{markdown}
# Your Role

You are an expert AI who specializes in improving the effectiveness of cross-modal text-image semantic similarity search from a vector database containing image embeddings computed by a multimodal CLIP model.

# Your Task

You will receive a user query and have to rewrite them into clear, specific, caption-like queries suitable for retrieving relevant images from the vector database.

Keep in mind that your rewritten query will be sent to a vector database, which does cross-modal similarity search for retrieving images.
\end{minted}
\end{tcolorbox}

\subsection{Text-to-Text Similarity Search}
\begin{tcolorbox}[
    enhanced, 
    breakable,
    skin first=enhanced,
    skin middle=enhanced,
    skin last=enhanced,
]
\begin{minted}[fontsize=\footnotesize,breaklines]{markdown}
# Your Role

You are an expert AI who specializes in improving the effectiveness of textual semantic similarity search from a vector database containing text embeddings.

# Your Task

You will receive a user query and have to rewrite them into clear, specific, and concise queries suitable for retrieving relevant information from the vector database.

Keep in mind that your rewritten query will be sent to a vector database, which does semantic similarity search for retrieving text.
\end{minted}
\end{tcolorbox}

\section{Image Analysis Prompts}
\label{appendix:image_prompts}
In the following we provide the system instructions for image analysis functionalities within \collex.

\subsection{VQA System Instruction}
\begin{tcolorbox}[
    enhanced, 
    breakable,
    skin first=enhanced,
    skin middle=enhanced,
    skin last=enhanced,
]
\begin{minted}[fontsize=\footnotesize,breaklines]{markdown}
# Your Role

You are an expert AI assistant that specializes in performing accurate Visual Question Answering (VQA) on images.

# Your Task

You will receive a question, an image, and metadata about the image from a user.
Then you must generate an accurate but concise answer to that question based on the image and the metadata.
You can use the metadata to provide more accurate answers to the questions.
If a question cannot be answered based on the image (and metadata) alone, you can ask the user for additional information.
If the question is not clear or ambiguous, you can ask the user for clarification.
Keep in mind that the question can be about any aspect of the image, and your answer must be relevant to the question.
Do not hallucinate or provide incorrect information; only answer the question based on the image and metadata.
\end{minted}
\end{tcolorbox}
\subsection{Image Captioning System Instruction}
\begin{tcolorbox}[
    enhanced, 
    breakable,
    skin first=enhanced,
    skin middle=enhanced,
    skin last=enhanced,
]
\begin{minted}[fontsize=\footnotesize,breaklines]{markdown}
# Your Role

You are an expert AI assistant that specializes in performing accurate Image Captioning on images.

# Your Task

You will receive an image and additional metadata from a user and must generate a detailed and informative caption for that image.
The caption should describe the image in detail, including any objects, actions, or scenes depicted in the image.
You can use any available metadata about the image to generate a more accurate and detailed caption.

Keep in mind that the caption must be informative and descriptive, providing a clear understanding of the image to the user.
Do not provide generic or irrelevant captions; focus on the content and context of the image.
If the user requires the caption to be concise, you can generate a shorter version of the caption.
\end{minted}
\end{tcolorbox}

\subsection{OCR System Instruction}
\begin{tcolorbox}[
    enhanced, 
    breakable,
    skin first=enhanced,
    skin middle=enhanced,
    skin last=enhanced,
]
\begin{minted}[fontsize=\footnotesize,breaklines]{markdown}
# Your Role

You are an expert AI assistant that specializes in performing accurate Optical Character Recognition on images.

# Your Task

You will receive an image and additional metadata from a user and must extract and recognize text from that image.
You should provide the user with the extracted text from the image, ensuring accuracy and completeness.
You can use any available metadata about the image to improve the accuracy of the text extraction.

Keep in mind that the extracted text must be accurate and complete, capturing all relevant information from the image.
Do not provide incorrect or incomplete text; ensure that the extracted text is as accurate as possible.
\end{minted}
\end{tcolorbox}

\subsection{Object Detection System Instruction}
\begin{tcolorbox}[
    enhanced, 
    breakable,
    skin first=enhanced,
    skin middle=enhanced,
    skin last=enhanced,
]
\begin{minted}[fontsize=\footnotesize,breaklines]{markdown}
# Your Role

You are an expert AI assistant that specializes in performing accurate Object Detection on images.

# Your Task

You will receive an image and additional metadata from a user and must identify and locate prominent objects within that image.
You should provide the user with a list of objects detected in the image including their detailed descriptions and approximate locations.
You can use any available metadata about the image to improve the accuracy of the object detection.
Keep in mind that the object detection results must be accurate and complete, identifying all relevant objects in the image.
Do not provide incorrect or incomplete object detection results; ensure that all objects are correctly identified and described.

# Output Format

Output all detected objects in JSON format with the following structure:
```json
[
    {
         "name": "<NAME OF THE OBJECT>",
         "description": "<DESCRIPTION OF THE OBJECT>",
         "bounding_box": {
             "x": 100,
             "y": 100,
             "width": 50,
             "height": 50
        }
    }
]
```
\end{minted}
\end{tcolorbox}

\newpage
\section{System Demonstration}
\label{appendix:user_stories}
In the following we provide high-resultion screenshots of the user stories from Section~\ref{sec:demo}.
\subsection{General Functionality}
\begin{figure*}[!ht]
    \centering
    \begin{subfigure}{0.8\linewidth}
        \includegraphics[width=\textwidth]{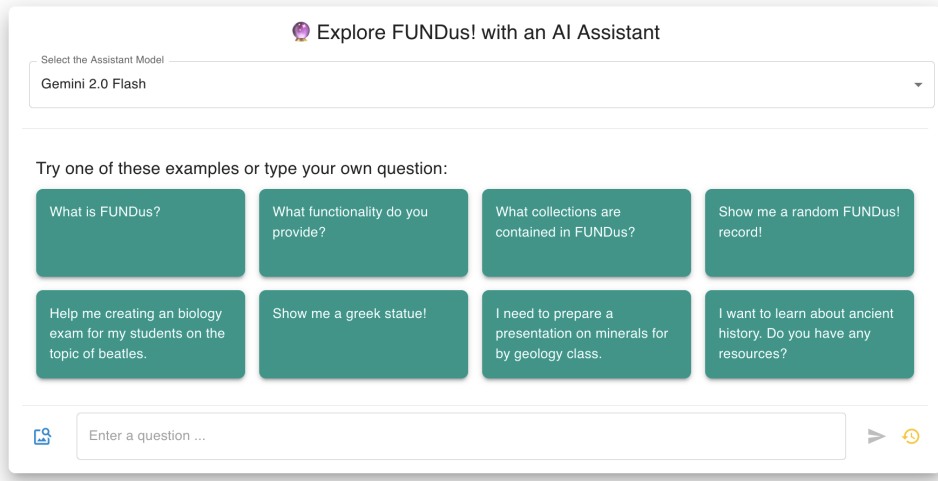}
        \subcaption{The \collex start page.}
        \label{fig:demo:highres:general:1}
    \end{subfigure}

    \begin{subfigure}{0.8\linewidth}
        \includegraphics[width=\textwidth]{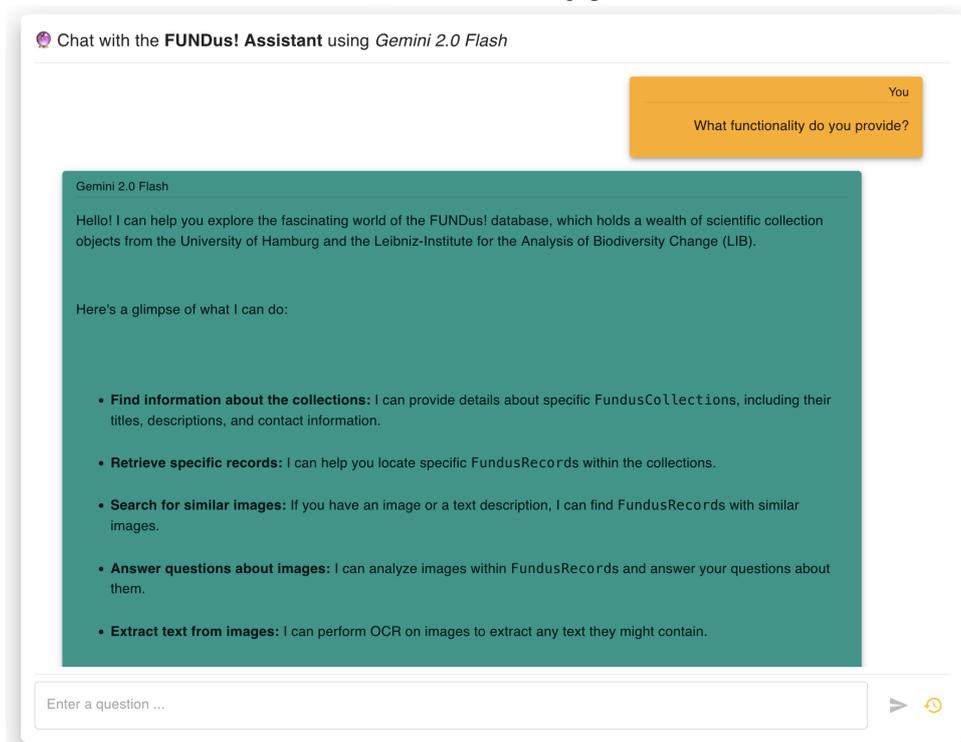}
        \subcaption{General functionality query.}
        \label{fig:demo:highres:general:2}
    \end{subfigure}
    
    \caption{Show-casing \collex general functionality.}
\end{figure*}
\begin{figure*}[!ht]\ContinuedFloat
    \centering
    \begin{subfigure}{0.8\linewidth}
        \includegraphics[width=\textwidth]{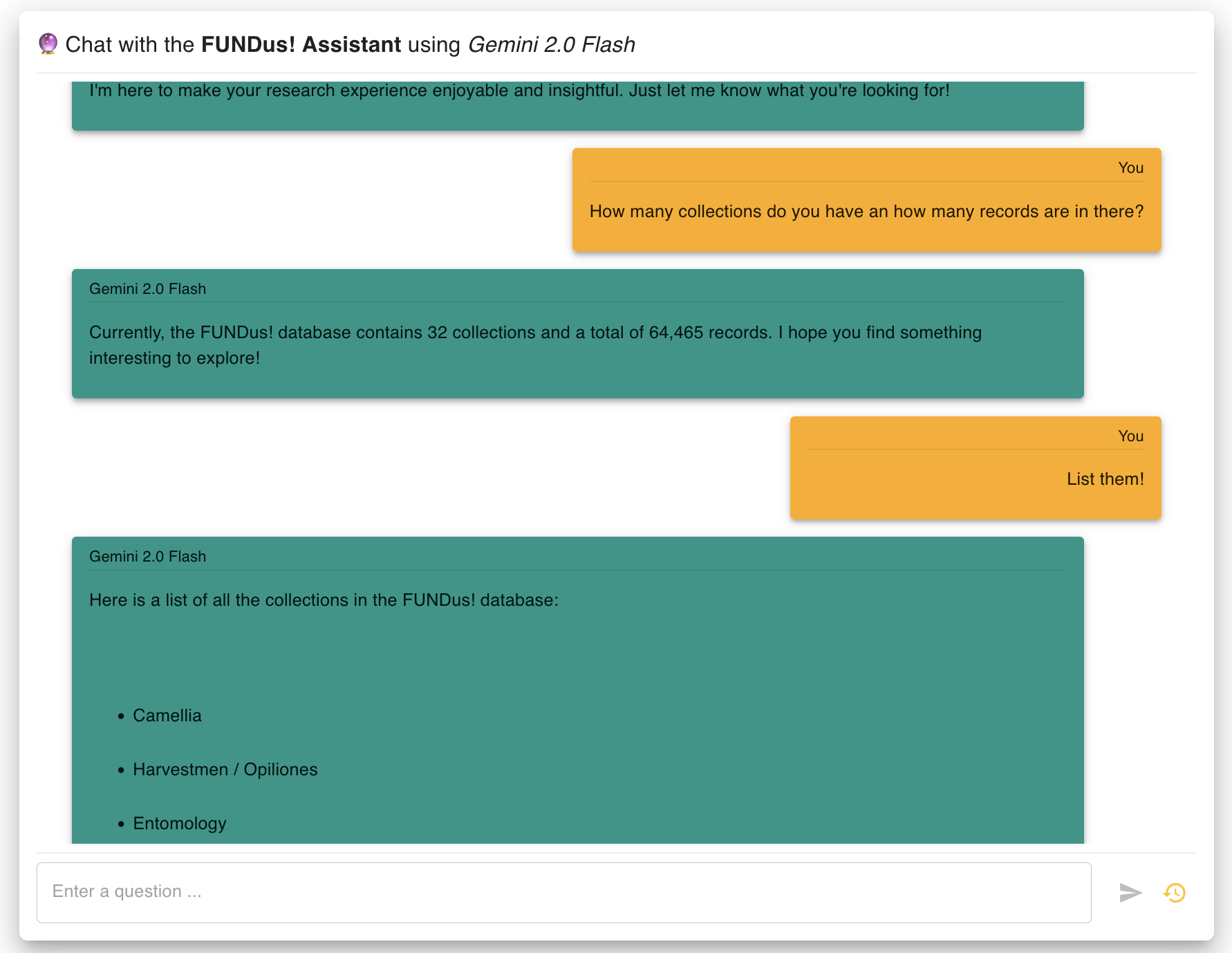}
        \subcaption{Records and Collections query.}
        \label{fig:demo:highres:general:3}
    \end{subfigure}
    
    \caption{Show-casing \collex general functionality.}
    \label{fig:demo:highres:general}
\end{figure*}

\FloatBarrier
\clearpage
\subsection{Geology Class Presentation}
\begin{figure*}[!ht]
    \centering
    \begin{subfigure}{0.8\linewidth}
        \includegraphics[width=\textwidth]{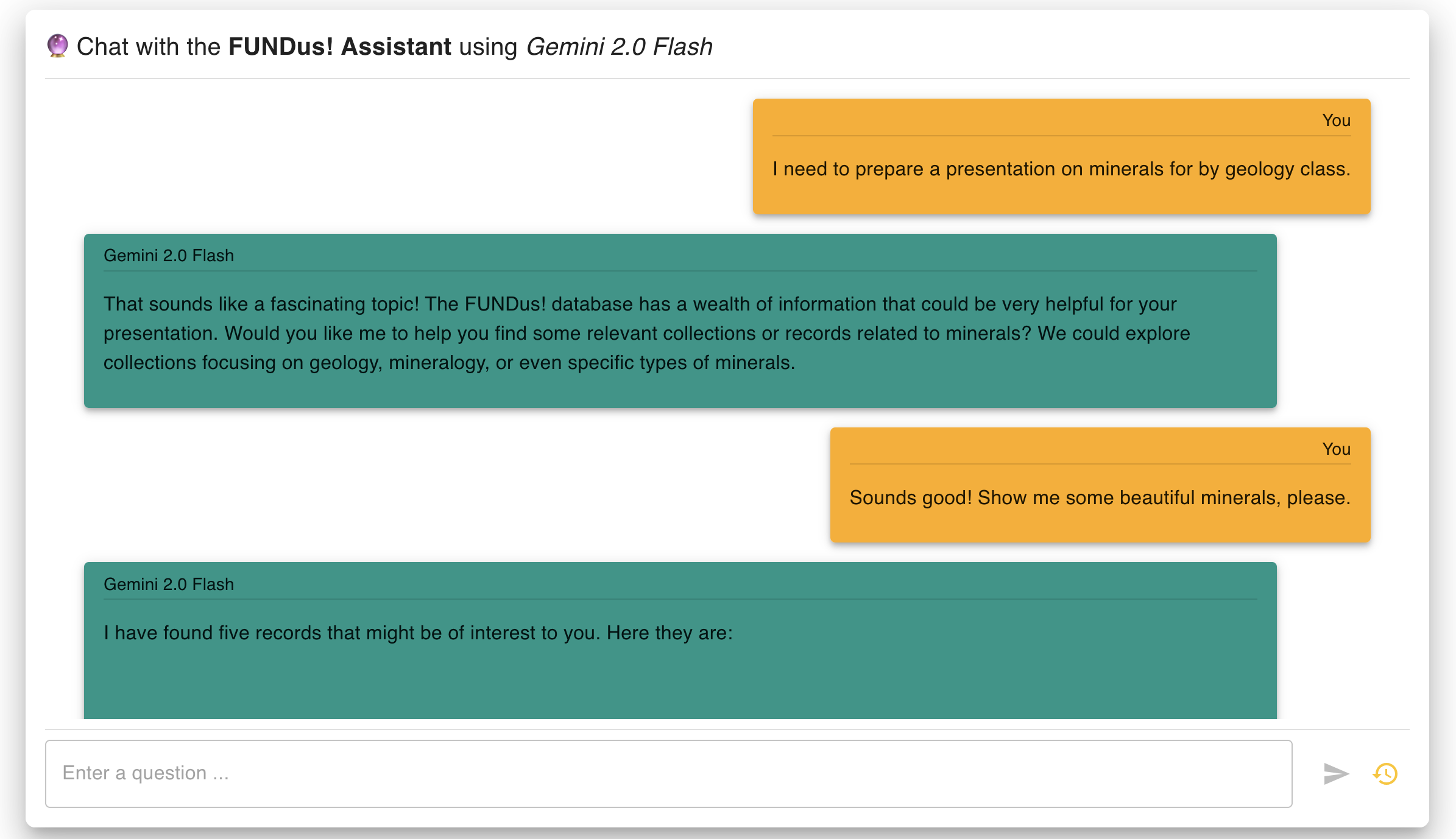}
        \subcaption{Start of the chat.}
        \label{fig:demo:highres:geo_class:1}
    \end{subfigure}

    \begin{subfigure}{0.8\linewidth}
        \includegraphics[width=\textwidth]{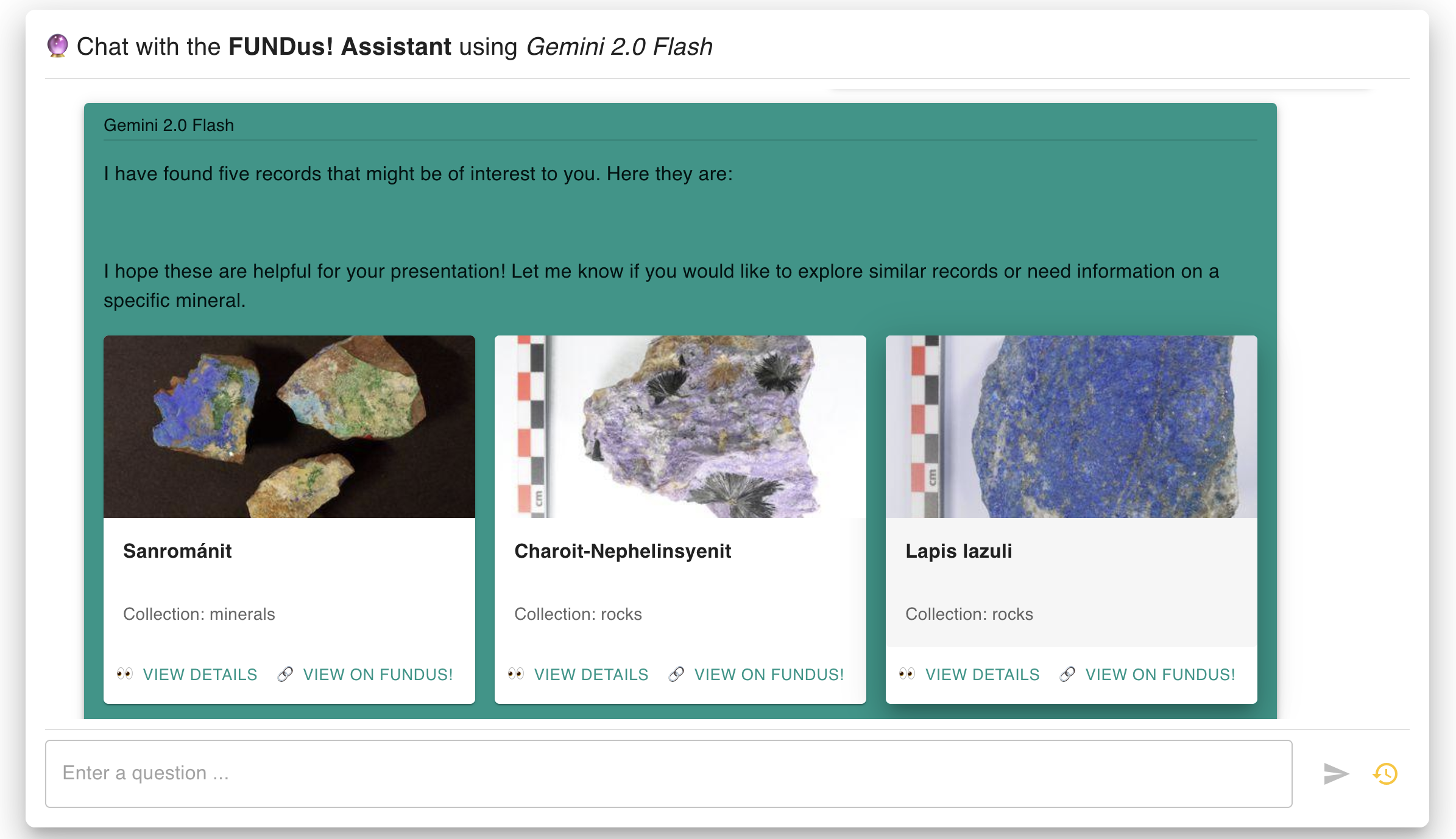}
        \subcaption{Search results for the user query.}
        \label{fig:demo:highres:geo_class:2}
    \end{subfigure}
    
    \caption{A demonstration of \collex based on an exemplary use case of getting inspiration for a geology class presentation.}
    \label{fig:demo:highres:geo_class}
\end{figure*}
\begin{figure*}[!ht]\ContinuedFloat
    \centering
    \begin{subfigure}{0.8\linewidth}
        \includegraphics[width=\textwidth]{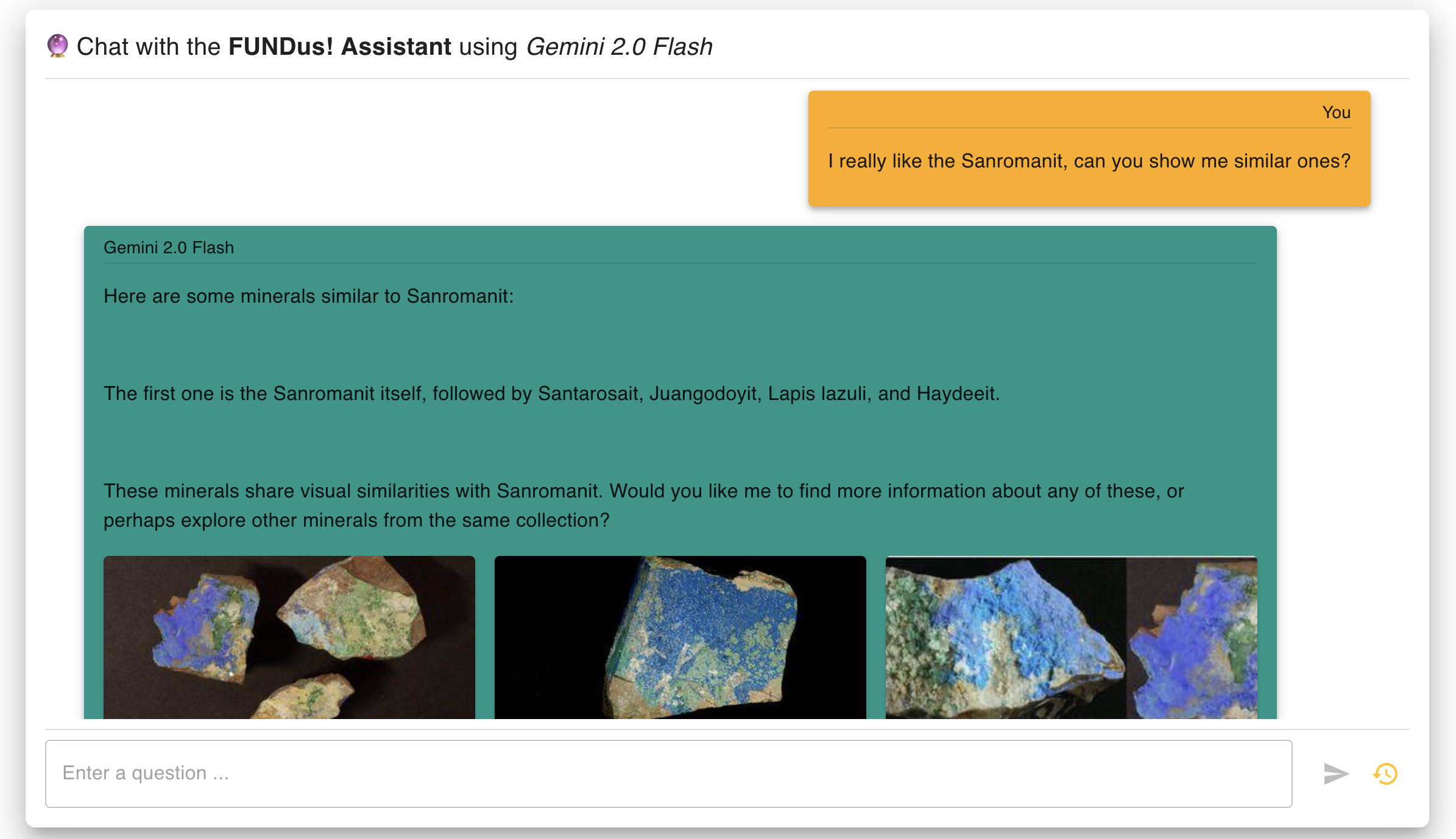}
        \subcaption{Image similarity search results.}
        \label{fig:demo:highres:geo_class:3}
    \end{subfigure}

    \begin{subfigure}{0.8\linewidth}
        \includegraphics[width=\textwidth]{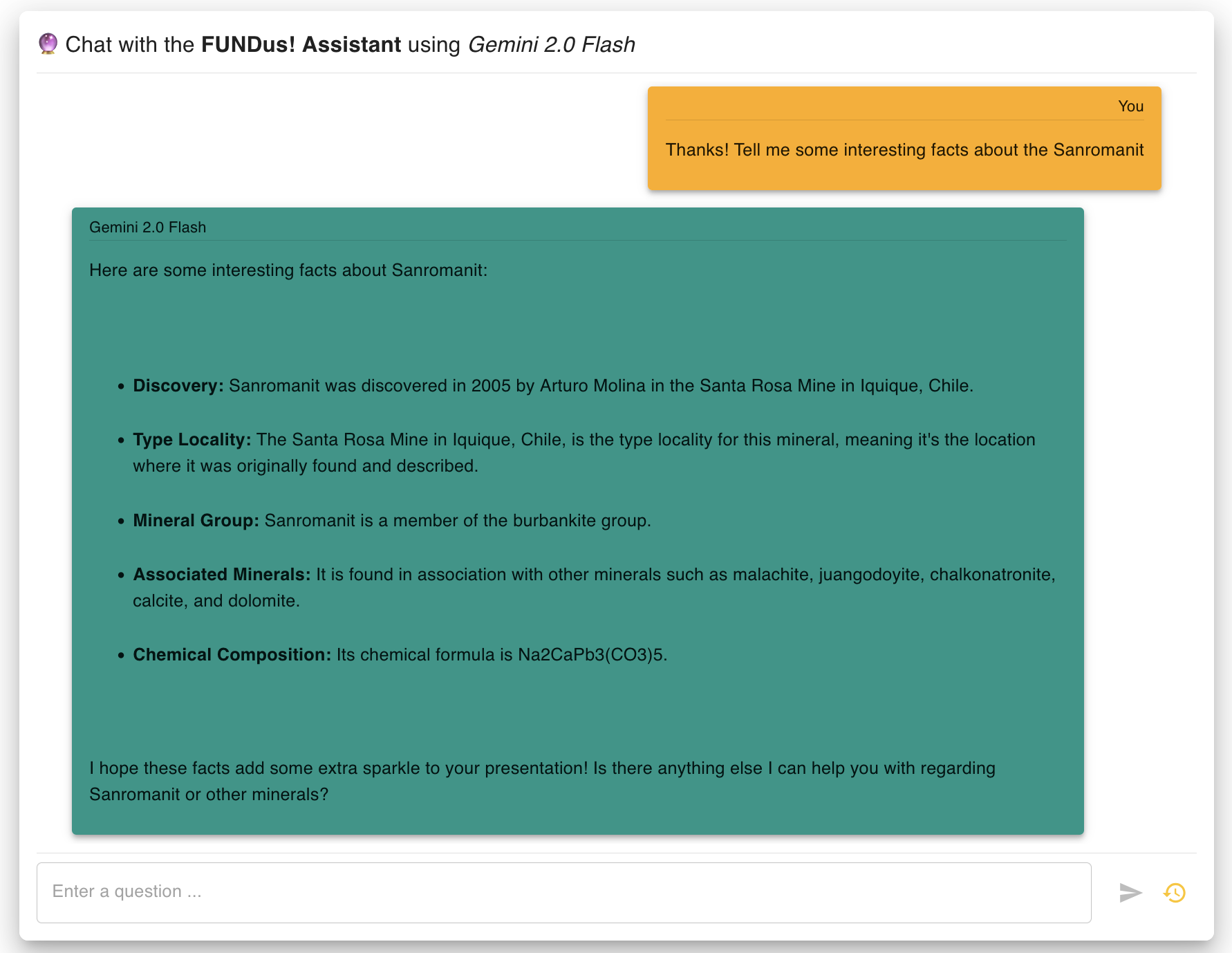}
        \subcaption{Requesting more details.}
        \label{fig:demo:highres:geo_class:4}
    \end{subfigure}
    
    \caption{A demonstration of \collex based on an exemplary use case of getting inspiration for a geology class presentation.}
\end{figure*}
\begin{figure*}[!ht]\ContinuedFloat
    \centering
    \begin{subfigure}{0.8\linewidth}
        \includegraphics[width=\textwidth]{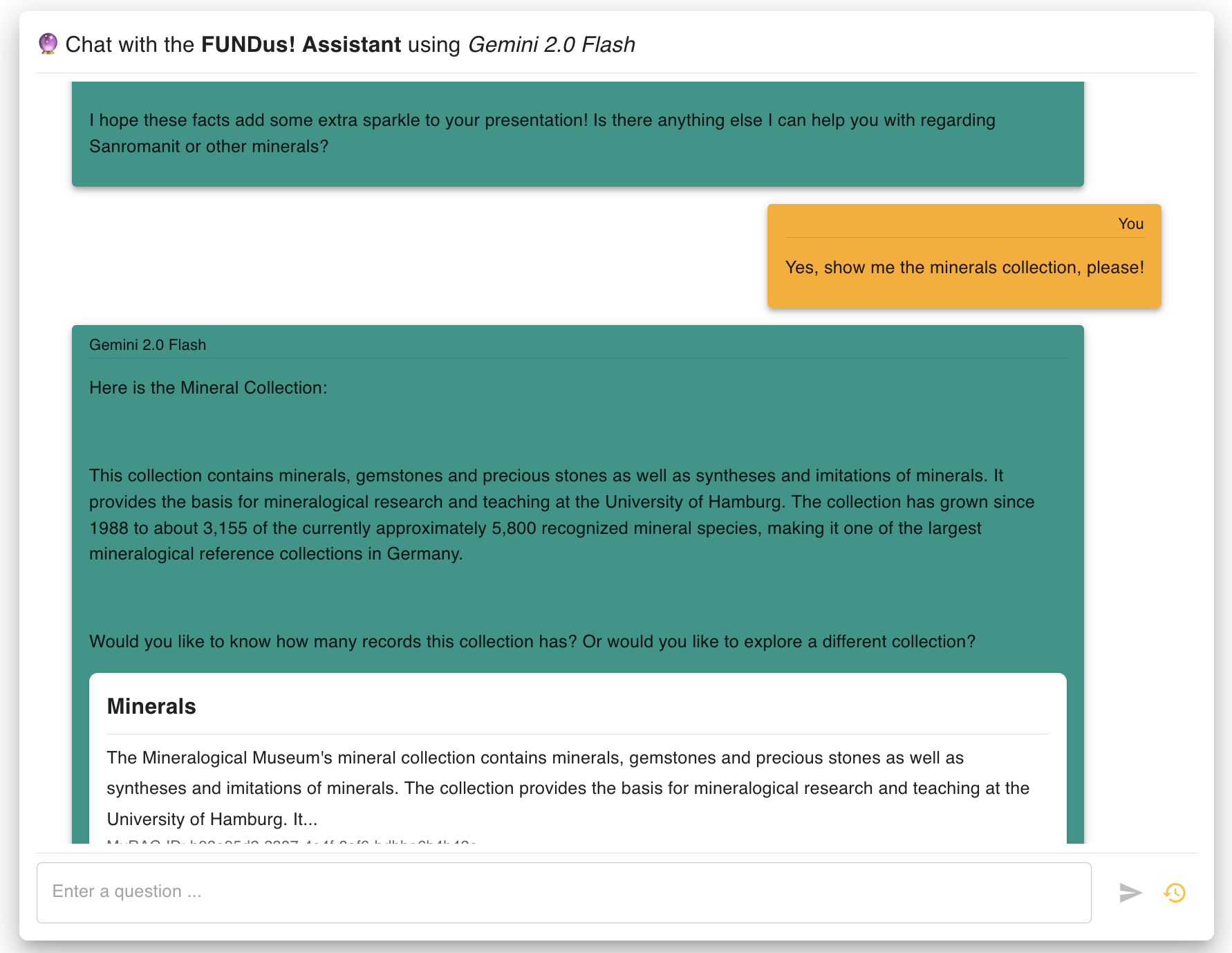}
        \subcaption{Showing the minerals collection.}
        \label{fig:demo:highres:geo_class:5}
    \end{subfigure}

    \begin{subfigure}{0.8\linewidth}
        \includegraphics[width=\textwidth]{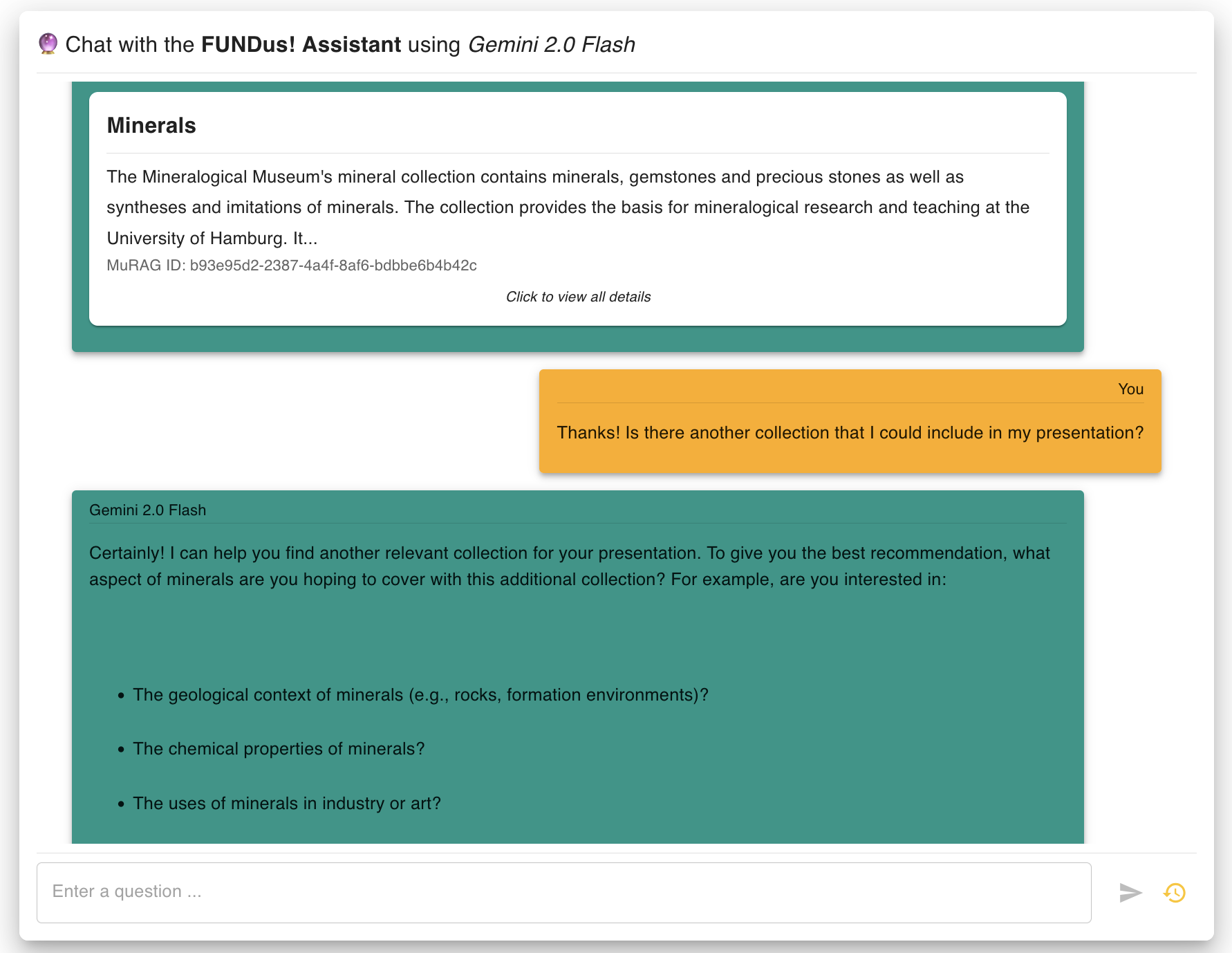}
        \subcaption{Follow-up query.}
        \label{fig:demo:highres:geo_class:6}
    \end{subfigure}
    
    \caption{A demonstration of \collex based on an exemplary use case of getting inspiration for a geology class presentation.}
\end{figure*}

\FloatBarrier
\subsection{Finding an Exhibition Piece}
\begin{figure*}[!ht]
    \centering
    \begin{subfigure}{0.8\linewidth}
        \includegraphics[width=\textwidth]{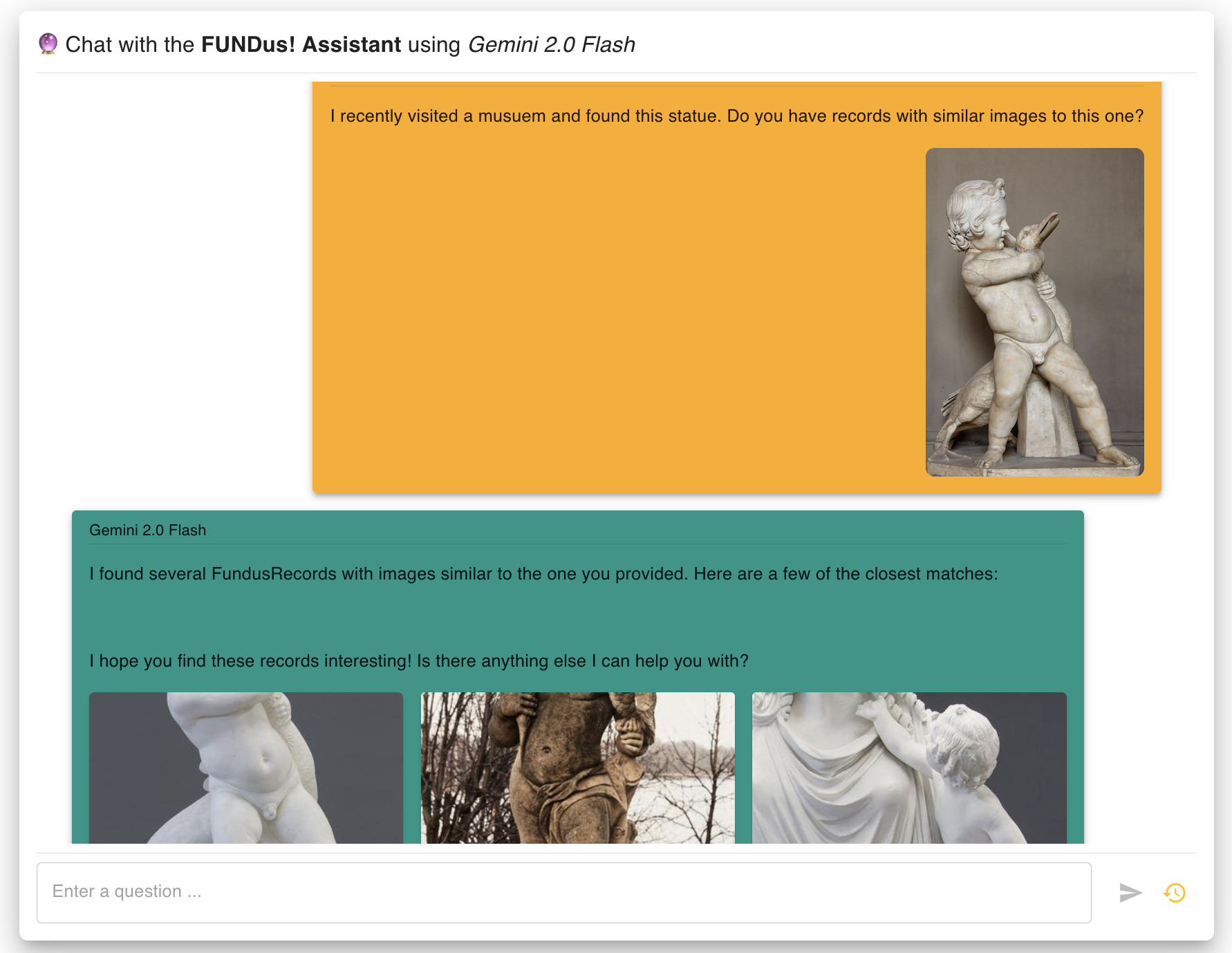}
        \caption{Text-image search request and results.}
        \label{fig:demo:highres:sim_ex:1}
    \end{subfigure}

    \begin{subfigure}{0.8\linewidth}
        \includegraphics[width=\textwidth]{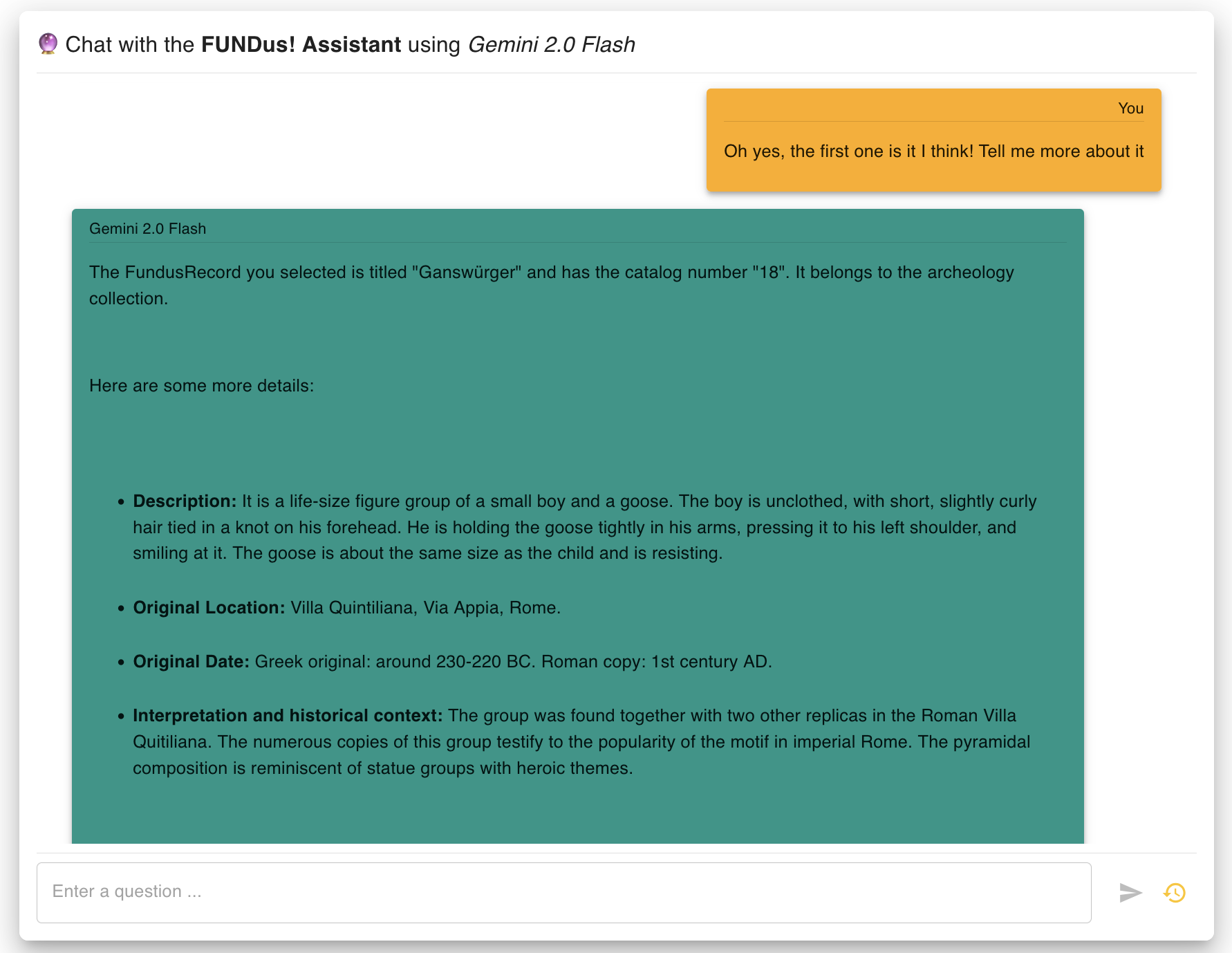}
        \caption{Follow-up details query.}
        \label{fig:demo:highres:sim_ex:2}
    \end{subfigure}
    
    \caption{A demonstration of \collex based on an exemplary use case of finding an exhibition piece.}
    \label{fig:demo:highres:sim_ex}
\end{figure*}
\begin{figure*}\ContinuedFloat
    \centering
    \begin{subfigure}{0.8\linewidth}
        \includegraphics[width=\textwidth]{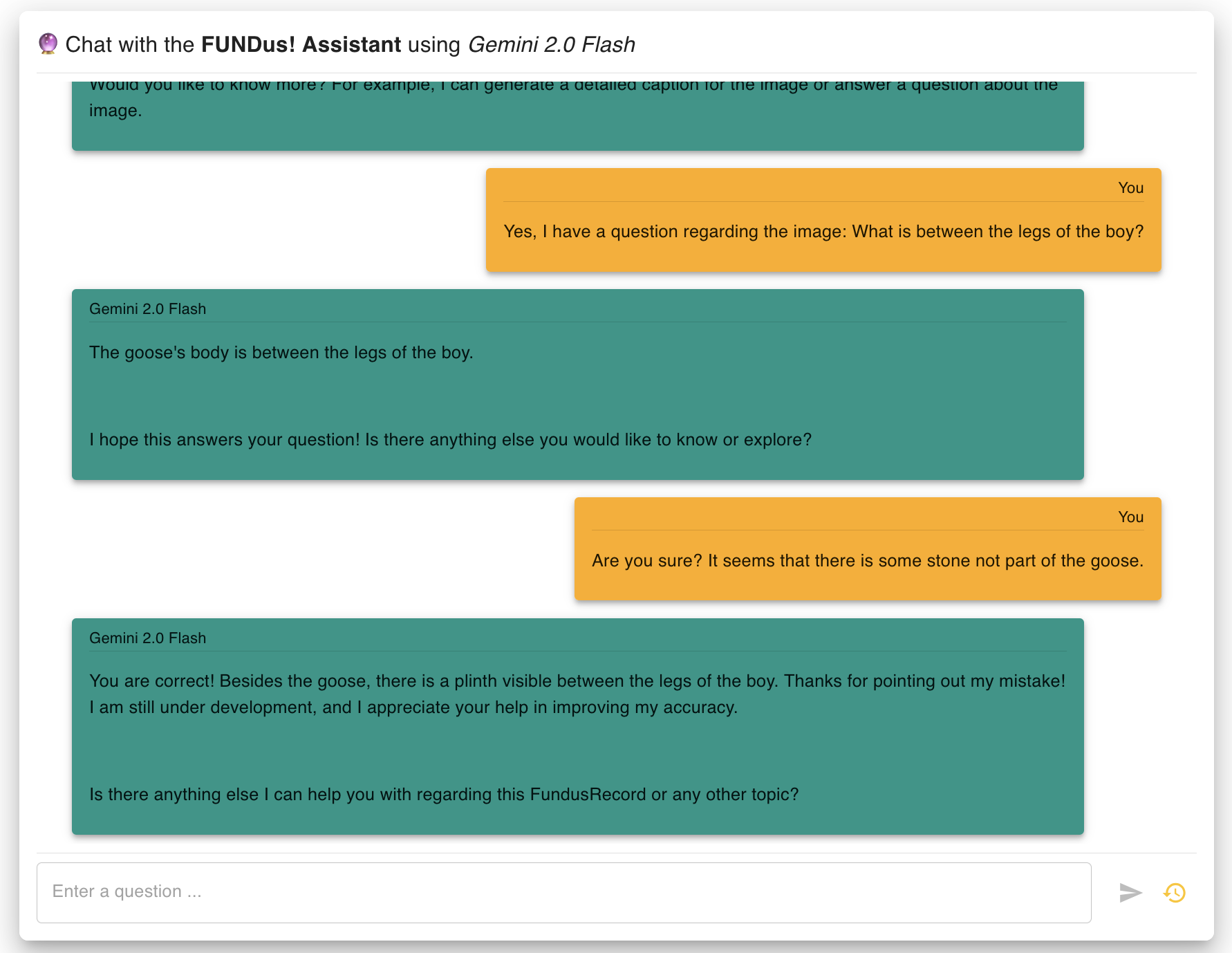}
        \caption{Image analysis queries.}
        \label{fig:demo:highres:sim_ex:3}
    \end{subfigure}

    \caption{A demonstration of \collex based on an exemplary use case of finding an exhibition piece.}
\end{figure*}

\end{document}